\documentclass[aps,prd,reprint,superscriptaddress,nofootinbib,floatfix]{revtex4-2}

\usepackage{graphicx}
\usepackage{dcolumn}
\usepackage{bm}

\usepackage{amsmath}
\usepackage{amssymb}
\usepackage{multirow}
\usepackage{tabularx}   
\usepackage{dcolumn}    
\usepackage{booktabs}   
\usepackage{siunitx}    
\newcolumntype{d}{D{.}{\times}{-1}} 
\usepackage{float}
\usepackage[caption=false]{subfig}
\usepackage[T1]{fontenc}
\usepackage[utf8]{inputenc}
\DeclareUnicodeCharacter{2009}{\thinspace}
\usepackage{orcidlink}
\usepackage{todonotes}
\usepackage{hyperref}

\hypersetup{colorlinks,linkcolor={blue},citecolor={blue},urlcolor={blue}}

\begin{document}

\preprint{APS/123-QED}

\title{Dynamical friction on black holes in scalar field environment}

\author{Adith Praveen}
  \email{ph21b001@smail.iitm.ac.in}
\author{Dawood Kothawala}%
 \email{dawood@iitm.ac.in}
\affiliation{Department of Physics, Indian Institute of Technology Madras, Chennai-600036, India}%

\date{\today}

\begin{abstract}
Astrophysical black holes exist within non-vacuum environments, and their motion through these environments generically result in a force on these black holes through processes such as dynamical friction and Bondi accretion. We explore these forces numerically for black holes moving through a complex scalar field of mass $m$ with constant acceleration $a$, while also revisiting the constant velocity case considered in the existing literature. The former is modeled by the C-metric, while we use Painlev\'e-Gullstrand metric with an additional divergence and vorticity free velocity field to mimic the constant velocity motion. Our simulations reveal several novel and interesting aspects of the drag force in both cases. For the constant velocity case, the force saturates at late times, with a value that depends on velocity in a manner distinct from known dependence. For the constant acceleration case, the force increases to a maximum value $F_\star$ and then reduces drastically, with $F_\star \approx 0.3~ m a$. Moreover, the density wake in this case shows revivals separated by decreasing time intervals.
\end{abstract}

\maketitle


\section{Introduction}
\label{Introduction}

Black holes (BHs) provide unique regions to study the evolution of matter fields under extreme conditions. Since astrophysical BHs do not live in vacuum conditions, the signals coming from them, both electromagnetic and gravitational, can carry imprints of the interaction between the BHs and their environment. For example, a binary BH merger occurring in a non-vacuum environment can have  external forces acting on it from the environment altering the trajectory from the vacuum case, thereby affecting the gravitational wave waveform. With the detection of gravitational waves from LIGO-VIRGO-KAGRA~\cite{gw}, the study of such imprints of environmental effects on BHs has become even more relevant from an observational point of view. 

As a BH moves through an environment, it experiences two kinds of forces, namely dynamical friction and Bondi accretion. Dynamical friction is a drag force which occurs when a massive body moves through a medium, forming a high energy density wake of the field behind it due to gravitational interaction. This wake then exerts a net force on the BH in the direction opposite to its motion. This effect is called dynamical friction. The very first work on such environmental effects on compact objects, revealing the key features of dynamical friction, was by Chandrasekhar~\cite{chandrasekhardynamical1943}, who obtained the dynamical friction force on a star moving in an environment of lighter point masses; to leading order,
\begin{equation}
    F_d \propto \frac{\log(v)}{v^2}.
\end{equation}
The above scaling breaks down in the limits $v \to 0$ and $v \to 1$. Subsequent works have added pressure and relativistic corrections to this expression~\cite{traykovadynamical2021}. In the case that we shall be interested in (discussed in next paragraph), it is not apriori clear what elements of Chandrasekhar's derivation would survive, which is what makes the analyses relevant. Similar comments apply to the Bondi accretion component of the force, resulting from accretion of field momenta into the BH. 

Specifically, our environment will be modeled by a complex scalar field, which has recently being considered~\cite{traykovadynamical2021} in the context of interaction of BHs with a dark matter environment, with dark matter composed of light scalar particles such as axions modeled by a complex scalar field following the Klein-Gordon equation. Be that as it may, we emphasize that we take complex scalar field as the environment solely as the simplest case beyond the point particle case. Our key focus is on analyzing how the modification of spacetime geometry due to BH motion affects the considerations of force on the BH vis-a-vis the existing work in the literature, where the motion is modeled through a coordinate transformation representing boost. Previous works have considered forces on BHs moving through a scalar field medium with a constant velocity~\cite{traykovadynamical2021,dysonrelativistic2024,wanggravitational2024}. These works start with Schwarzschild (or Kerr) BH in isotropic coordinates and then boost it to simulate a moving BH. We here focus on approaching the problem in a different manner, that does not involve a boost coordinate transformation. We consider two kinds of BH motion: (i) motion with constant acceleration, described by an exact vacuum solution, and (ii) a BH moving with constant velocity, described, in the spirit of analogue models, by a Painlev\'e-Gullstrand metric with a specific constant velocity field added to it (the resultant metric is not a vacuum solution).
%
Studying dynamical friction of accelerating BHs has not been done in literature, although, from an astrophysical point of view, accelerating BH are more likely to occur (for instance, during binary BH mergers) than BHs with constant velocity. Our results reveal several peculiar and interesting features that appear in this case, which we discuss in detail.

The paper is divided into four sections. Section \ref{Scalar Field} describes the basic equations involving the scalar field stress tensor within the ADM formalism. In section \ref{Background Spacetime}, we describe the spacetimes that we use in this work as a background for evolving the scalar field. Section \ref{Numerical Simulation and Analysis} discusses the code used for the analysis as well as the basic analytical results to compute the force acting on the BH. Sections \ref{Results} and \ref{Conclusion} gives the results and key conclusions that emerge from this work. We use Latin indices to denote spacetime indices (0,1,2,3) and Greek indices to denote space indices (1,2,3).

\section{Scalar Field}
\label{Scalar Field}

In this work, we consider a complex scalar field as the simplest model for the environment, and evolve it on the fixed background spacetime of the BH ignoring the backreaction (which is ensured by suitable choice of parameters). The scalar field $\varphi(x^i)$ is assumed to have mass $m$ and evolves according to the Klein-Gordon equation
\begin{equation}
    (\nabla^i \nabla_i  - m^2) \varphi = 0,
\end{equation}
Using the ADM formalism, we can split this equation into two first order equations in time given by
\begin{widetext}
\begin{align}
    \partial_t \varphi &= \alpha \Pi + \beta^\mu \partial_\mu \varphi, \\
    \partial_t \Pi &= \alpha \gamma^{\mu\nu}\partial_\mu\partial_\nu \varphi + \alpha(\text{Tr}(\mathbf{K}) \Pi - \gamma^{\mu\nu}\Gamma^\delta_{\mu\nu} \partial_\delta\varphi - m^2 \varphi) + \partial_\mu\varphi \partial^\mu \alpha + \beta^\mu\partial_\mu\Pi,
\end{align}
\end{widetext}
where $\Pi$ is the conjugate momentum of $\varphi$, $\alpha$ is lapse, $\beta^\mu$ is shift, $\Gamma^\delta_{\mu\nu}$ is the three Christoffel symbol, $\gamma^{\mu\nu}$ is the three metric and $\text{Tr}(\mathbf{K})$ is the trace of extrinsic curvature. The stress-energy tensor for this field is given as,
\begin{equation}
    T_{ij} = \nabla_{(i} \varphi \nabla_{j)} \varphi^* - \frac{1}{2}g_{ij} (\nabla^k \varphi \nabla_k \varphi^* + \frac{1}{2} \mu^2 |\varphi|^2).
\end{equation}
Finally, to ensure that we can ignore backreaction, we impose the condition
\begin{equation}
    \rho M^2 \ll 1
\end{equation}
where $M$ is the mass of the BH and $\rho = T_{ij}n^i n^j$ is the energy density of the scalar field.

\section{Background Spacetimes}
\label{Background Spacetime}

As mentioned in the introduction, we shall be mainly interested in how the distortion in spacetime geometry due to relative motion between the BH and its environment affects the forces acting on it. In particular, we shall consider the following two cases:
\begin{enumerate}
    \item Vacuum solution describing accelerating BH, represent by the so called C-metric whose Riemann curvature explicitly depends on acceleration. 
    \item Schwarzschild BH in the Painleve-Gullstrand gauge, altered by addition of a divergence and vorticity free velocity field to represent relative motion through the environment. This physically motivated metric happens to be a non-vacuum solution, whose source has an elegant form which we briefly comment upon.
\end{enumerate}

\subsection{\label{sec:level2}Accelerating BHs}

As mentioned, we shall use the C metric solution to study accelerating BHs. The C-metric satisfies the vacuum Einstein equations, and there are several subtleties and issued related to the interpretation of this metric~\cite{griffithsinterpreting2006}. However, we will here use it the regime where these subtleties do not affect the conclusions.  

One of the useful forms of the C-metric is~\cite{KinnCM},
\begin{equation}
    ds^2 = \frac{1}{A^2 (x+y)^2} \left( -F dt^2 + \frac{dy^2}{F} + \frac{dx^2}{G} + G d\phi^2 \right),
\end{equation}
where $G$ and $F$ are given as,
\begin{equation}
\begin{split}
    G &= 1 - x^2 - 2 \tilde{M} A x^3, \\
    F &= -1 + y^2 - 2 \tilde{M} A y^3, \\
\end{split}
\end{equation}

We will assume that $27 \tilde{M}^2 A^2< 1$, in which case both $G$ and $F$ possess three distinct real roots. In our work, we will use the form of the C-metric introduced by Hong and Teo~\cite{hongnew2003}, which offers a considerable simplification over the old form. They used the linear transformation in $x$ and $y$ along with a rescaling of $t$, $\phi$, $\tilde{M}$ and $A$ in such a way that the root structure of the cubic equation is expressed in a simple way. These transformations fix the roots of the metric function in $x$ at $+1$, $-1$ and $-1/(2a M)$. It is also necessary to assume that $0<2a M<1$ so as to preserve the order of the roots. Their final expression for the metric element of the C-metric is,
\begin{equation}
    ds^2 = \frac{1}{a^2 (x+y)^2} \left( -F d\tau^2 + \frac{dy^2}{F} + \frac{dx^2}{G} + G d\varphi^2 \right),
    \label{xyform2}
\end{equation}
where,
\begin{align}
    G &= (1-x^2)(1 + 2 a M x), \\
    F &= -(1-y^2)(1 - 2 a M y).
\end{align}
For the signature of the metric to remain Lorentzian, a necessary condition is $G>0$. To enforce this, $x$ has to lie between appropriate roots of $G$. Furthermore, there exists a conformal infinity at $x+y=0$. Considering all this, we restrict ourselves to $x \in (-1,1)$ and $x+y>0$. In this domain, the metric represents a pair of accelerated BHs \cite{griffithsinterpreting2006}. 

For better interpretation of results, the C-metric in \eqref{xyform2} is transformed to spherical polar coordinates. Using the restrictions in $x$ and $y$ imposed, we will use the following transformation given in~\cite{griffithsinterpreting2006},
\begin{align}
    x &= \cos\theta, \\
    y &= \frac{1}{a r},\\
    \tau &= a t_c,
\end{align}
with $\theta \in (0,\pi)$. In these new coordinates the metric takes the form,
\begin{equation}
\begin{split}
    ds^2 = \frac{1}{(1 + a r \cos\theta)^2}\Bigg( -Q dt_c^2 + \frac{dr^2}{Q} + \frac{r^2 d\theta^2}{P} \\ +P r^2 \sin^2\theta \text{ }d\phi^2 \Bigg),
    \\~~
    \\~~
\end{split}
\label{cmsph}
\end{equation}
where,
\begin{align}
    P &= 1 + 2 a M \cos\theta, \\
    Q &= (1-a^2 r^2) \left(1 - \frac{2 M }{r}\right),
\end{align}

and it reduces to Schwarzschild when $a = 0$, where $a$ is the acceleration of the BH~\cite{griffithsinterpreting2006}.

In the form given in equation \eqref{cmsph}, the C-metric has two Killing horizons; at $r=2 M$ and $r=1/a$, which dictates the domain of our simulation. We consider values of $a$ such that $a M \ll 1$ to push the $r=1/a$ Killing horizon sufficiently far away. Furthermore, it is an asymptotically flat Type D spacetime~\cite{drayasymptotic1982,ashtekarexistence1981}, which we exploit for the force calculation in the upcoming section. The C-metric also has a conical singularity, which comes through the the range of $\varphi \in (-\pi C, \pi C)$, where $C$ is another parameter of the metric \cite{griffithsinterpreting2006}.

For simulation purposes, the form of the C-metric in equation \eqref{cmsph} is not ideal for computational purposes due to the shift vector being zero. To mitigate this, another coordinate transformation is performed, which is as follows
\begin{equation}
    t = t_c + \int \sqrt{\frac{2 m}{r}}\frac{1}{1 - \frac{2 m}{r}} dr.
\end{equation}
The transformation performed is the Painlev\`e-Gullstrand transformation of Schwarzschild metric. This further helps in interpretation of the results by using to the river model understanding associated with PG metric~\cite{hamiltonriver2008}. The metric in our choice of coordinates is given as,
\begin{widetext}
\begin{equation}
\begin{split}
    ds^2 &= \frac{1}{(1 + a r \cos\theta)^2} \Bigg[ -\left(1 - \frac{2 M}{r}\right)(1 - a^2 r^2) dt^2 + 2 \sqrt{\frac{2 M}{r}}(1 - a^2 r^2) dt dr +\\ 
    &\quad \frac{r - 2 M (1 - a^2 r^2)}{(r - 2 M)(1 - a^2 r^2)} dr^2 + \frac{r^2}{1+2 a M \cos\theta} d\theta^2 + \frac{r^2 (1+2 a M \cos\theta)}{(1 + 2 a M)^2} d\phi^2 \Bigg].
\end{split}
\end{equation}
\end{widetext}

\subsection{Painlev\`e-Gullstrand  metric}
\label{Painleve-Gullstrand  metric}

For simulating a BH moving with constant velocity, we use a modified version of the Painlev\`e-Gullstrand(PG) metric for Schwarzschild BH. In this section we describe this modified metric. The general PG line element is of the form~\cite{fischerspace-time2003},

\begin{equation}
    ds^2 = -(1 - \vec{v}^2) dt^2 - 2 \vec{v}.d\vec{r} + d\vec{r}^2,
    \label{eq:base}
\end{equation}

where $\vec{v}$ is a three vector which can be interpreted as the velocity of the flow of spacetime. This interpretation can be taken in the form of river model of spacetime~\cite{hamiltonriver2008}, where the flow is given by the three vector $\vec{v}$. 

To get a stationary BH, put $\vec{v} = - \sqrt{\frac{2 M}{r}} \hat{r}$, which gives the Schwarzschild metric in PG coordinates. This is given as,

\begin{equation}
    ds^2 = -\left(1 - \frac{2 M}{r} \right)dt^2 + 2 \sqrt{\frac{2 M}{r}} dt dr + dr^2 + r^2 d\Omega^2.
\end{equation}

To get a moving BH, we add to the Schwarzschild velocity vector a constant velocity in the $z$ direction, $\vec{v_0} = -v_0 \hat{z}$. We thus have $\vec{v} = - \sqrt{\frac{2 M}{r}} \hat{r} -v_0 \hat{z}$. With this, the metric looks as,
\begin{widetext}
\begin{equation}
\begin{split}
    ds^2 = -\left(1 - \frac{2 M}{r} - v_0^2 -\sqrt{\frac{2 M}{r}} v_0 \cos\theta  \right) dt^2 + 2 \left(\sqrt{\frac{2 M}{r}} +  v_0 \cos\theta  \right)dt dr \\+ 2 v_0 r sin\theta  dt d\theta+ dr^2 + r^2 d\Omega^2.
\end{split}
\end{equation}
\end{widetext}

Unlike the C-metric, this metric is not a vacuum solution and has a non-zero Einstein tensor. Since we are dealing with a PG metric, the formalism presented in \cite{fischerspace-time2003} can be used to express the Einstein tensor in terms of the velocity vector of the PG metric. This is done to get a better interpretation of the non-zero stress tensor. For the discussion that follows we will use the Cartesian coordinate system for the spatial coordinates unless specified. We start with first defining a tetrad frame for the PG metric as,
\begin{equation}
\begin{aligned}
    e^{\hat{t}}_{~t} &= 1, \qquad & e^{\hat{t}}_{~\mu} &= 0, \\
    e^{\hat{\mu}}_{~t} &= -v^\mu, \qquad & e^{\hat{\mu}}_{~\nu} &= \delta^{\hat{\mu}}_{~\nu}.
    \label{eq:tetrad}
\end{aligned}
\end{equation}
where $v^\mu$ is the velocity vector mentioned in \eqref{eq:base}.

The inverse of the basis is,
\begin{equation}
\begin{aligned}
    e_{\hat{t}}^{~t} &= 1, \qquad & e_{\hat{t}}^{~\mu} &= v^\mu, \\
    e_{\hat{\nu}}^{~t} &= 0, \qquad & e_{\hat{\nu}}^{~\mu} &= \delta_{\hat{\nu}}^{~\mu}.
\end{aligned}
\end{equation}

In this tetrad basis, we can write the Einstein tensor for any PG metric as \cite{fischerspace-time2003},
\begin{equation}
\begin{aligned}
    G_{\hat{t}\hat{t}} &= \frac{1}{2}(\text{Tr}\mathbf{K})^2 - \frac{1}{2}\text{Tr}(\mathbf{K^2}), \\
    G_{\hat{t}\hat{\mu}} &= -\frac{1}{2}(\nabla\times\vec{\omega})_\mu,\\
    G_{\hat{\mu}\hat{\nu}} &= \frac{\text{d}}{\text{d}t}(K_{\mu\nu} - \delta_{\mu\nu}\text{Tr}\mathbf{K}) + \text{Tr}\mathbf{K} \left(K_{\mu\nu} - \frac{1}{2} \delta_{\mu\nu}\text{Tr}\mathbf{K}\right) \\ &- \frac{1}{2} \delta_{\mu\nu} \text{Tr}(\mathbf{K^2}) - (\mathbf{K}\mathbf{\Omega} + \mathbf{\Omega}\mathbf{K})_{\mu\nu},
\end{aligned}
\end{equation}
where 
\begin{equation}
\begin{aligned}
K_{\mu\nu} &= \frac{1}{2}(\partial_\mu v_\nu + \partial_\nu v_\mu), \\
    \Omega_{\mu\nu} &= \frac{1}{2}(\partial_\mu v_\nu - \partial_\nu v_\mu), \\
    \vec{\omega} &= \nabla \times \vec{v},\\
    \text{Tr}\mathbf{K} &= \text{div}(\vec{v}).
\end{aligned}
\end{equation}



In our case,  $\vec{v} = -\sqrt{\frac{2 M}{r}}\hat{r} - v_0 \hat{z}$. Since $v_0$ is a constant, $K_{\mu\nu}$ becomes the extrinsic curvature for the Schwarzschild spacetime. From this, on simplification we get
\begin{equation}
\begin{aligned}
    G_{\hat{t}\hat{t}} &= 0, \\
    G_{\hat{t}\hat{\mu}} &= 0,\\
    G_{\hat{\mu}\hat{\nu}} &= \frac{\text{d} S_{\mu\nu}}{\text{dt}} + \partial_\delta v^\delta  S_{\mu\nu},\\
    ~~~
\end{aligned}
\end{equation}
where $S_{\mu\nu} = K_{\mu\nu} - \delta_{\mu\nu} \text{Tr}\mathbf{K}$. We can further simplify $G_{\hat{\mu}\hat{\nu}}$ to get the following
\begin{equation}
G_{\hat{\mu}\hat{\nu}} = -v_0 \frac{\partial}{\partial z} S_{\mu\nu}.
\end{equation}

Converting the full Einstein tensor in the tetrad basis to spherical polar coordinates, we get the following,
\begin{widetext}
\begin{equation}
G_{ij} = 
\begin{bmatrix}
0 & 0 & 0 & 0 \\
0 & 3 \sqrt{\frac{2 M}{r^{5}}} v_0 \cos{\theta} & \frac{3}{2} \sqrt{\frac{2 M}{r^{3}}} v_0 \sin{\theta} & 0 \\
0 & \frac{3}{2} \sqrt{\frac{2 M}{r^{3}}} v_0 \sin{\theta} & \frac{3}{4} \sqrt{\frac{2 M}{r}} v_0 \cos{\theta} & 0 \\
0 & 0 & 0 & \frac{3}{4} \sqrt{\frac{2 M}{r}} v_0 \cos{\theta} \sin{\theta}^2 \\
\end{bmatrix}
\end{equation}
\end{widetext}




In contrast to C-metric, the PG metric uses horizon penetrating coordinates. As a result, solving the scalar field equation does not require the excision of the horizon. However, it is still important to excise the physical singularity at $r=0$. Therefore, we choose our excision region as a sphere with $r= 1.0 M$. 

For calculation of $\rho$ and force, we exclude a region slightly bigger than the Killing horizon. This is to ensure that $\rho>0$. For this metric the Killing horizon is given by the solution to the following equation,
\begin{equation}
1 - \frac{2 M}{r} - v_0^2 -\sqrt{\frac{2 M}{r}} v_0 \cos\theta = 0.
\end{equation} 
Since the Killing horizon has a $v_0$ dependence, the region excised for $\rho$ and force calculation varies with $v_0$. For example, for $v_0 = 0.8$, the excision region is a sphere with $r = 70.0 M$.

\section{Numerical Simulation and Analysis}
\label{Numerical Simulation and Analysis}

\subsection{Numerical Setup}

For the numerical simulation, we are using GRDzhadzha~\cite{Aurrekoetxea:2023fhl,Andrade2021}. This code allows us to model the scalar field on a background spacetime, under the assumption that the back-reaction on the spacetime from the field is negligible. 

We use the $4^{\text{th}}$ order finite differencing scheme for the derivatives and Runge-Kutta $4^{\text{th}}$ order for time integration. Furthermore, GRDzhadzha has adaptive mesh refinement (AMR) from Chombo which we have configured to be based on the background spacetime. This allows us to have more points near the horizon where the changes in the field values will occur over much smaller length scales. 

We have used $128^3$-point grid with 10 levels of AMR. The scalar field mass is set to $m M = 0.1$ for C-metric and $m M = 0.01$ for the PG metric. The grid length along each Cartesian axis is set as $L = 2048 M$ for C-Metric and $L = 4096 M$ for PG metric. The boundary conditions are set at extrapolating at zeroth order, which implies that the values at the edges of the grid are set to the same value as the nearest point in the bulk~\cite{traykovadynamical2021}. 

For the initial condition we have set the field $\varphi(t = 0,x,y,z) = 0.0001$ and the conjugate momentum $\Pi(t = 0,x,y,z) = 0.0001$. 

\subsection{Analysis : Force on the black hole}

In this section we describe our prescription to calculate the force on the BH based on the simulations. In our scenario, we simulate the evolution of the scalar field on a constant background spacetime, neglecting back-reaction from the scalar field. We justify this choice by initializing the scalar field to have a very low energy density, $\rho M^2 \ll 1$.

Since the spacetimes used in this work are asymptotically flat, we can use the following expression for the force along the $\mu^{\rm th}$ spacial axis,
\begin{equation}
    F_\mu = \frac{\text{d}P_\mu^{ADM}}{\text{d}t}
\end{equation}
where $P^{ADM}_\mu$ is the ADM momentum along the $\mu^{\rm th}$ spacial axis, and the last equality is made assuming the integrating surface is at spatial infinity. Since in simulations we have finite spacial regimes, the last equality will be an approximation. 

In this framework, we evaluate the first order response of the spacetime due to the scalar field by using its stress-energy tensor. For deriving the exact expression of the force, consider the current $J^i = T^i_j \xi^j$, where $\xi^i$ is not Killing vector of the entire spacetime. Now we use the following equality,
\begin{equation}
    \int \nabla_i J^i \sqrt{-g}\text{ } d^4x =  \int T^i_j \nabla_i \xi^j \sqrt{-g}\text{ } d^4x
\end{equation}
where the last equality comes from $\nabla_i T^i_j = 0$. Now we can use Gauss's law to simplify the equation into the following~\cite{cloughcontinuity2021}
\begin{equation}
\begin{split}
\frac{\partial}{\partial t}\left( \int_{\Sigma-\Sigma_{BH}} d^3 x\text{ } \sqrt{\gamma} Q  \right) =  -\int_{\partial\Sigma_{BH}} d^2 x \sqrt{\sigma} \text{ } F \\-\int_{\Sigma-\Sigma_{BH}} d^3 x \sqrt{\gamma}\text{ } S
\label{eq_integ_main}
\end{split}
\end{equation}
where $\Sigma$ is the 3D hypersurface on which we are doing the integration, $\Sigma_{BH}$ is the excised BH region, $\gamma$ and $\sigma$ are the determinants of the induced metrics along the 3D hypersurface and 2D surface $\partial\Sigma_{BH}$, and we define the other quantities as (note that we have used the ADM slicing in this setup), 
\begin{align}
    Q &= -n_i J^i \\
    F &= -\alpha N_\mu J^\mu \\
    S &= \alpha T^i_j \nabla_i \xi^j\\
\end{align}
where $N^\mu$ is the normal to surface $\partial\Sigma_{BH}$. Here we choose $\xi^i = \delta^i_\mu$, giving us,
\begin{equation}
\begin{split}
\frac{\partial}{\partial t}\left( \int_{\Sigma-\Sigma_{BH}} d^3 x \sqrt{\gamma}\text{ }  \alpha T^0_\mu \right) &=  \int_{\partial\Sigma_{BH}} d^2 x \sqrt{\sigma} \text{ } \alpha N_\nu T^\nu_\mu \\ &- \int_{\Sigma-\Sigma_{BH}} d^3 x \sqrt{\gamma} \text{ } \alpha T^i_j {}^{(4)}\Gamma^j_{i \mu}.\\
~~~
\end{split}
\end{equation}
The term on the LHS is the time derivative of the spatial integral of the momentum density of the field, which is the force acting on the BH. We can rewrite the second term to get the final expression for the force as
\begin{equation}
    F_\mu = -\int_{\partial\Sigma_{BH}} d^2 x \, \alpha \, \sqrt{\sigma} \, T^\nu_\mu \, N_\nu \,  - \int_{\Sigma-\Sigma_{BH}} d^3 x \,\alpha \,\sqrt{\gamma} \,  T^i_j \, {}^{(4)}\Gamma^j_{i \mu}.
\end{equation}
 Furthermore, we can integrate equation \eqref{eq_integ_main} with respect to time to get the following expression,
\begin{widetext}
\begin{equation}
\begin{split}
    \int_{\Sigma-\Sigma_{BH}} d^3 x\text{ } \sqrt{\gamma} Q \mid_{t=t_1} - \int_{\Sigma-\Sigma_{BH}} d^3 x\text{ } \sqrt{\gamma} Q \mid_{t=t_0} &=  -\int_{t=t_0}^{t=t_1}dt\int_{\partial\Sigma_{BH}} d^2 x \sqrt{\sigma} \text{ } F\\& -\int_{t=t_0}^{t=t_1}dt\int_{\Sigma-\Sigma_{BH}} d^3 x \sqrt{\gamma}\text{ } S
\end{split}
\end{equation}
\end{widetext}

which we use as a check for the numerical stability of the system, especially at late times. We only consider time points up to which the above identity holds.

\section{Results}
\label{Results}

In this section we present the results from our simulations.

\subsection{Accelerating BHs}

\begin{figure*}[htbp]
    \centering
    \subfloat[Initial formation of scalar field cloud.\label{fig:cmetric_cloud1}]{%
        \includegraphics[width=0.48\textwidth]{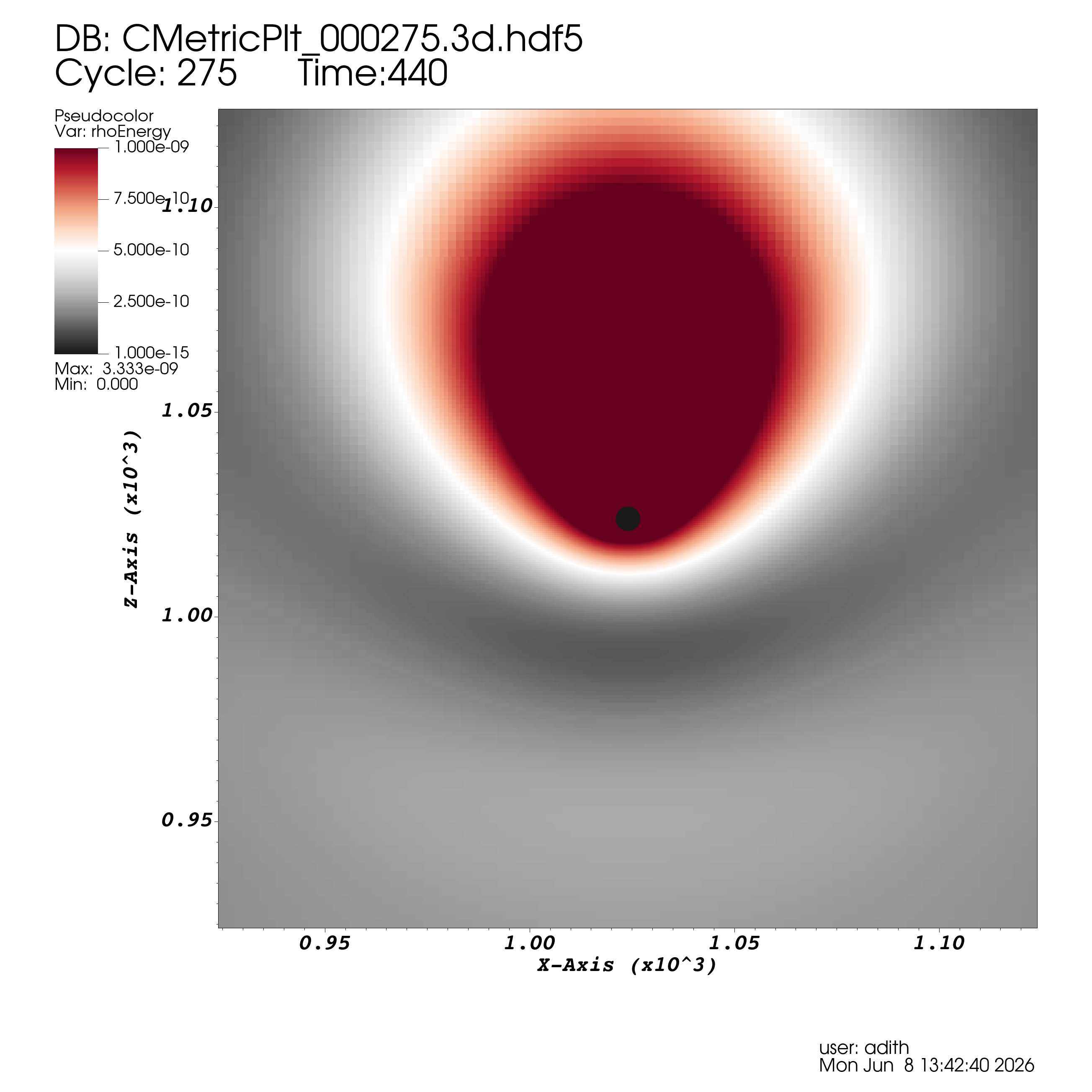}%
    }
    \hfill
    \subfloat[Detachment of the scalar field cloud.\label{fig:cmetric_cloud2}]{%
        \includegraphics[width=0.48\textwidth]{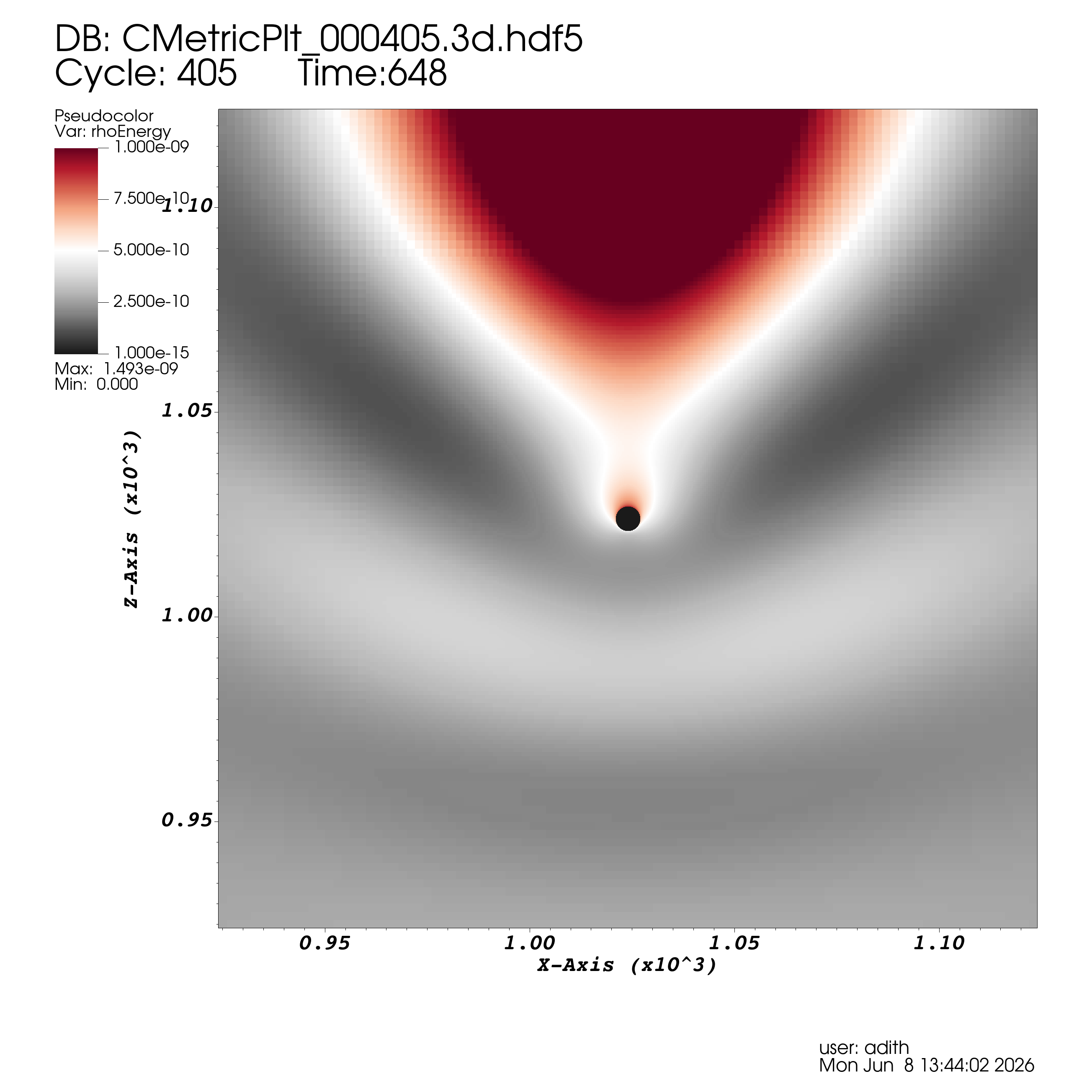}%
    }
    \vspace{0.5cm} 
    \subfloat[Formation of another scalar field cloud.\label{fig:cmetric_force_time}]{%
        \includegraphics[width=0.48\textwidth]{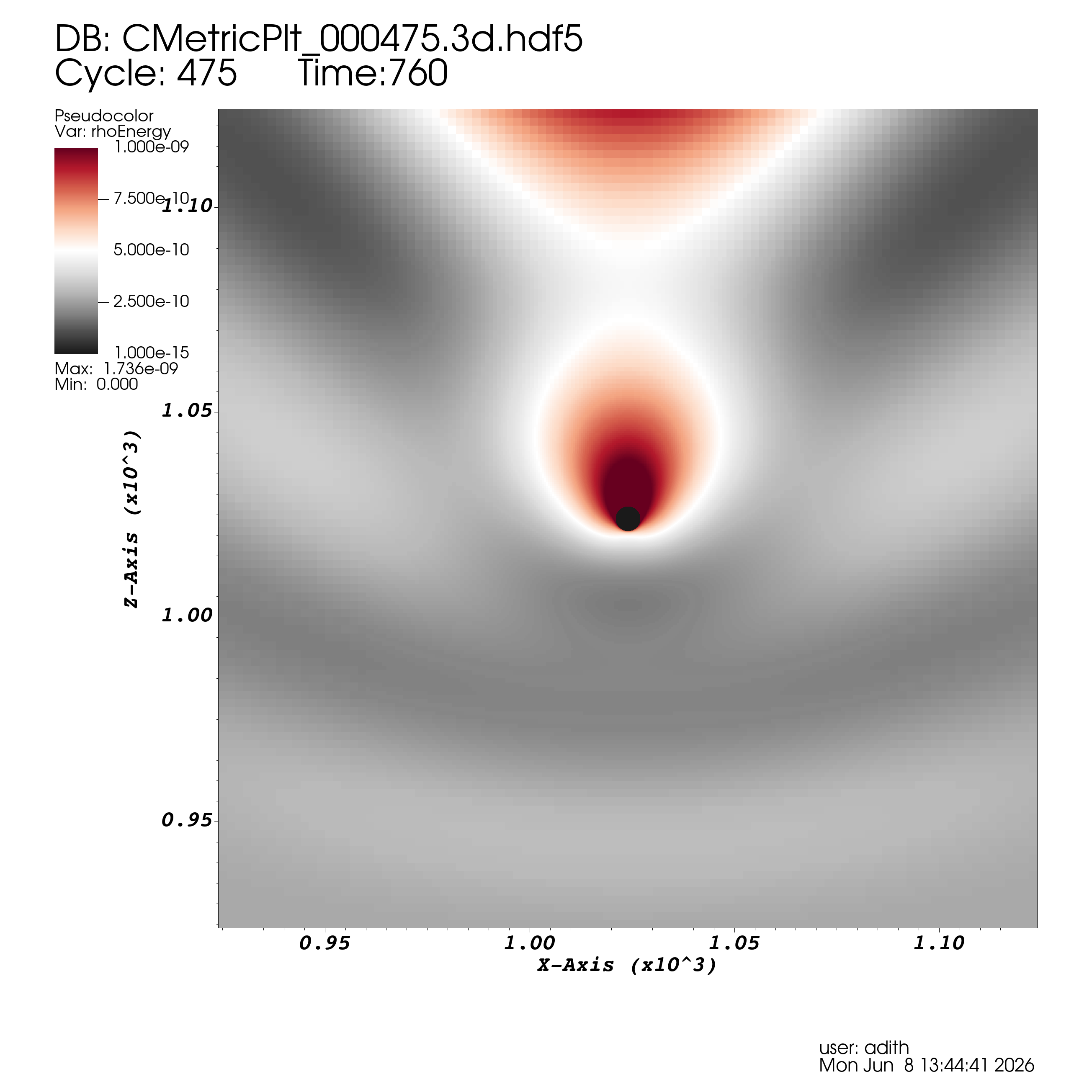}%
    }
    \hfill
    \subfloat[Late time characteristic of energy density.\label{fig:cmetric_force_alpha}]{%
        \includegraphics[width=0.48\textwidth]{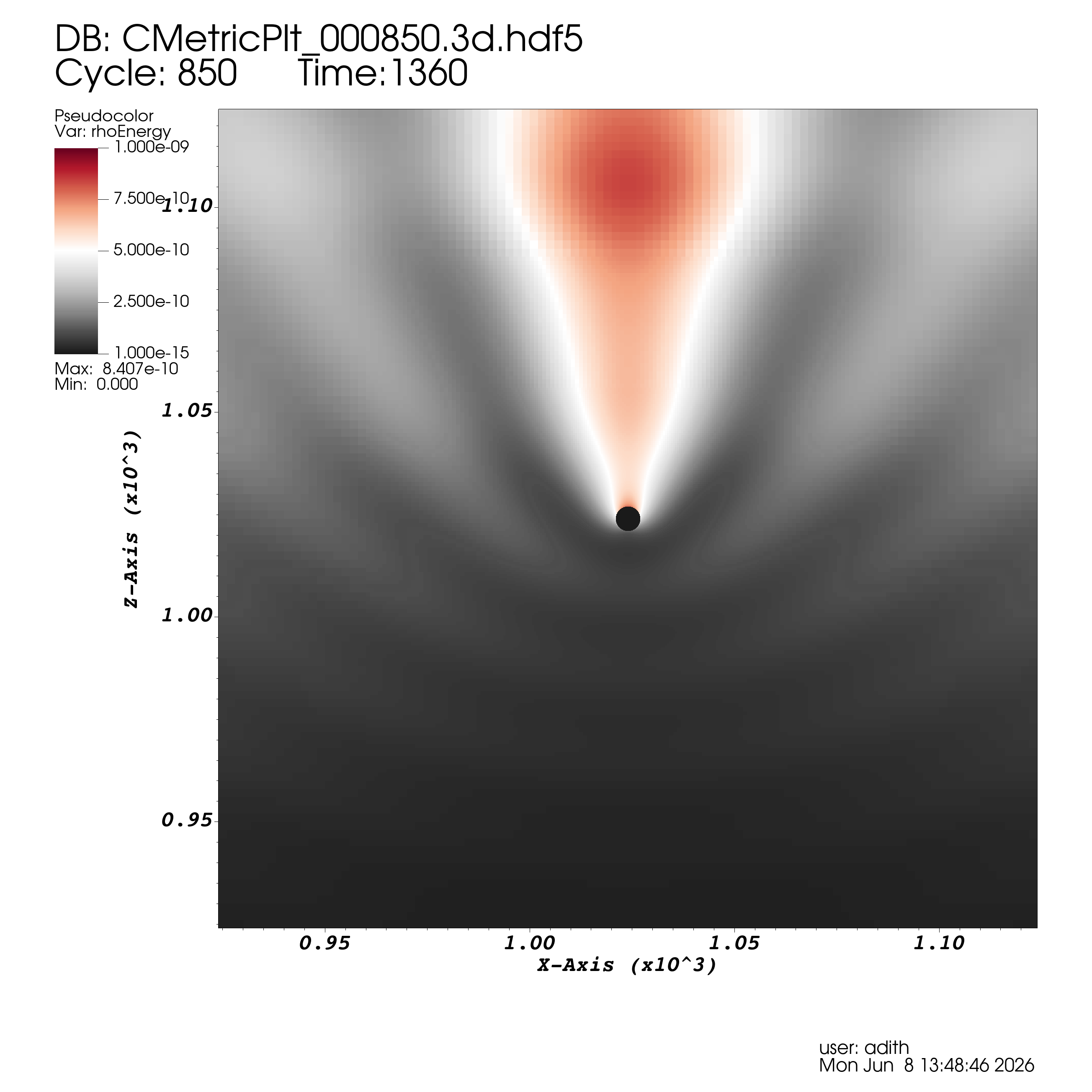}%
    }
    \caption{Evolution of the scalar field energy density with time for the C-metric case with $a = 0.0008$. The intensity and size of the scalar field energy density cloud increases initially, reaches a maxima and then reduces.}
    \label{fig:c_metric_results}
\end{figure*}

\begin{figure}[htbp]
    \centering
    \includegraphics[width=1\linewidth]{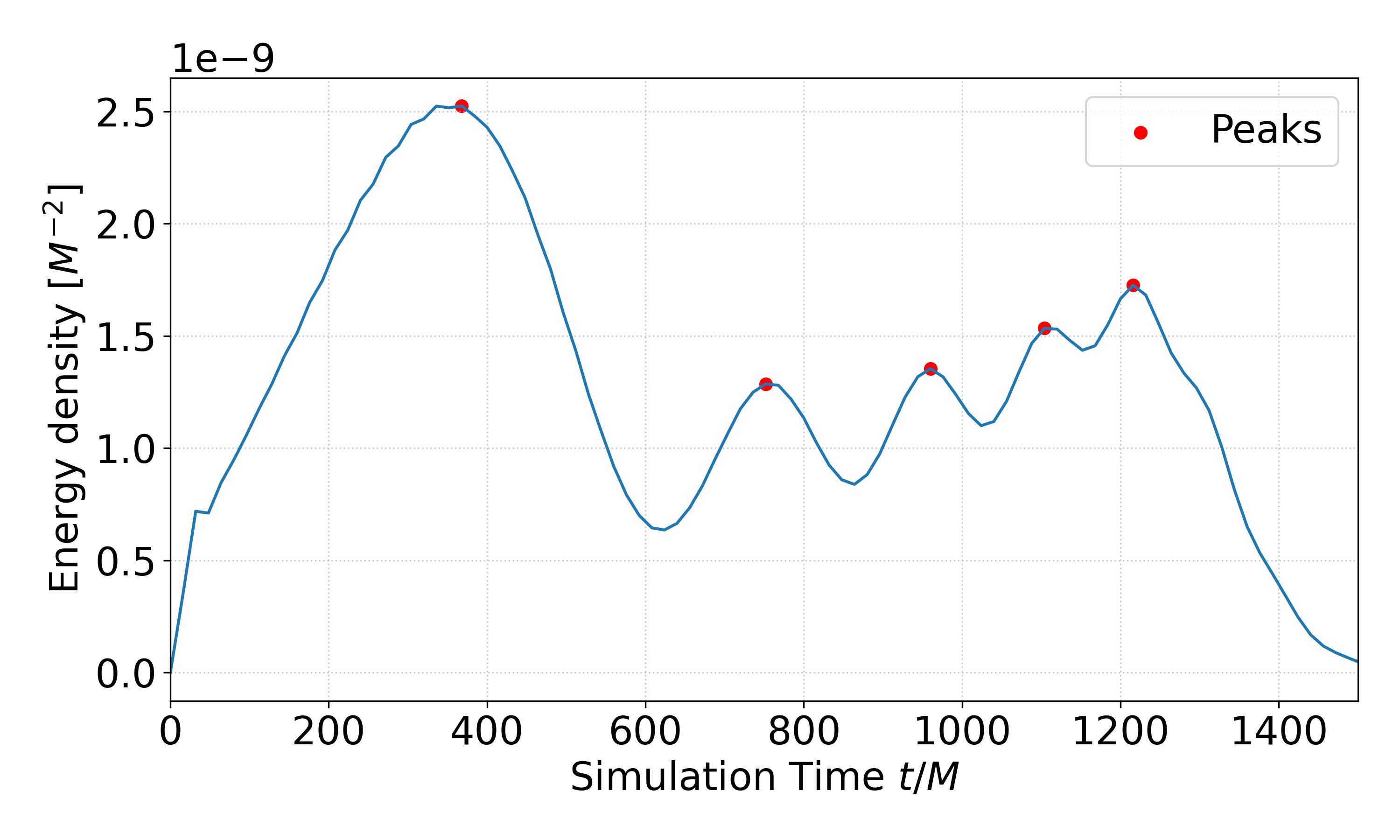}
    \caption{Variation of energy density versus time for the C-metric with $a = 0.0008$. This is for a point at a distance of $r=5M$ from the center of the black hole along the $z$-axis.}
    \label{fig:energy_density}
\end{figure}

It can be shown that the C-metric represents a BH accelerating in the negative $z$ direction~\cite{carneiro2022black}. Thereby, if dynamical friction occurs, it should lead to an increased energy density formation in the positive $z$ direction. In our simulations we observe this energy density formation and hence conclude that dynamical friction does occur for accelerating BHs. But the nature of the energy density distribution and force are different from the constant velocity case(described in \cite{traykovadynamical2021} and in the next section). There is an initial accumulation of a cloud of high energy density behind the motion of the BH, but then the cloud leaves the BH. Another cloud forms and leaves and this repeats for a few cycles, with the size of each cloud reducing with time. This feature can be seen in figures \ref{fig:c_metric_results} and \ref{fig:energy_density}. This characteristic of evolution is in sharp contrast to the constant velocity case where the energy density accumulation does not lead to clouds leaving the BH, and the features largely remain the same at late times. 

\begin{figure}[htbp]
    \centering
    \includegraphics[width=1\linewidth]{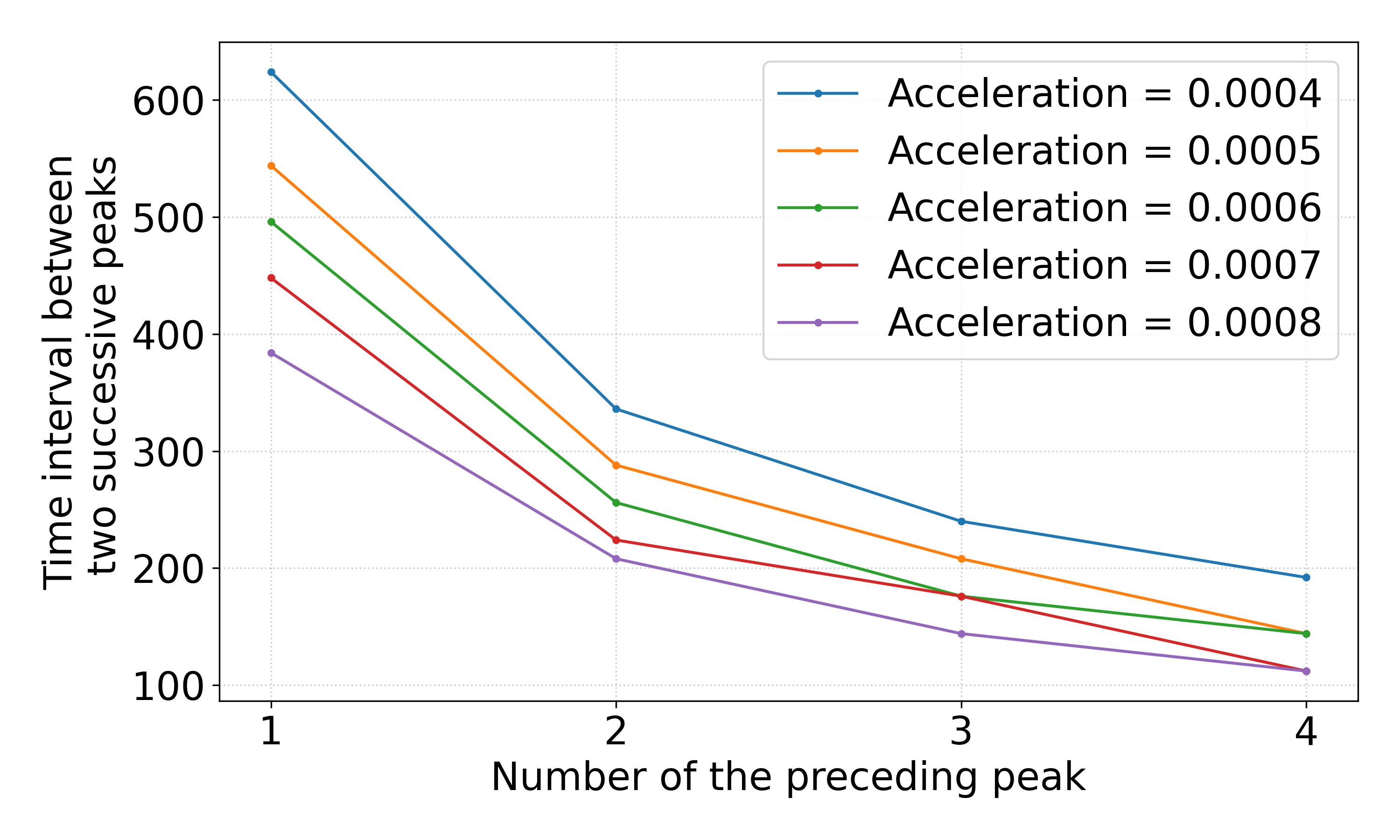}
    \caption{Variation of time period between the detachment of a scalar field cloud and formation of the next cloud, with simulation time, for the C-metric.}
    \label{fig:time_period_vs_t}
\end{figure}

\begin{figure}[htbp]
    \centering
    \includegraphics[width=1\linewidth]{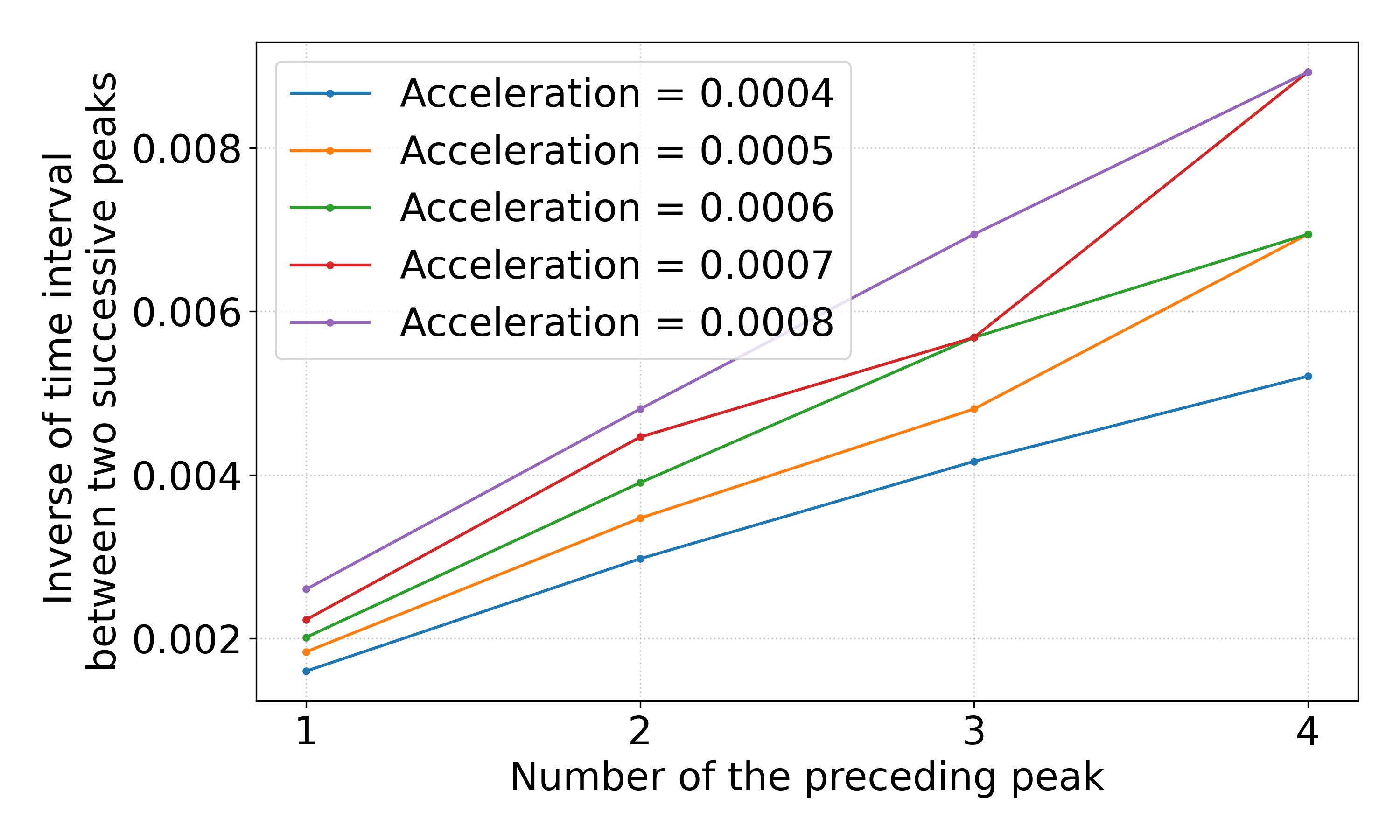}
    \caption{Variation of the inverse of time period between the detachment of a scalar field cloud and formation of the next cloud, with simulation time , for the C-metric.}
    \label{fig:inv_time_period}
\end{figure}

\begin{figure}[htbp]
    \centering
    \includegraphics[width=1\linewidth]{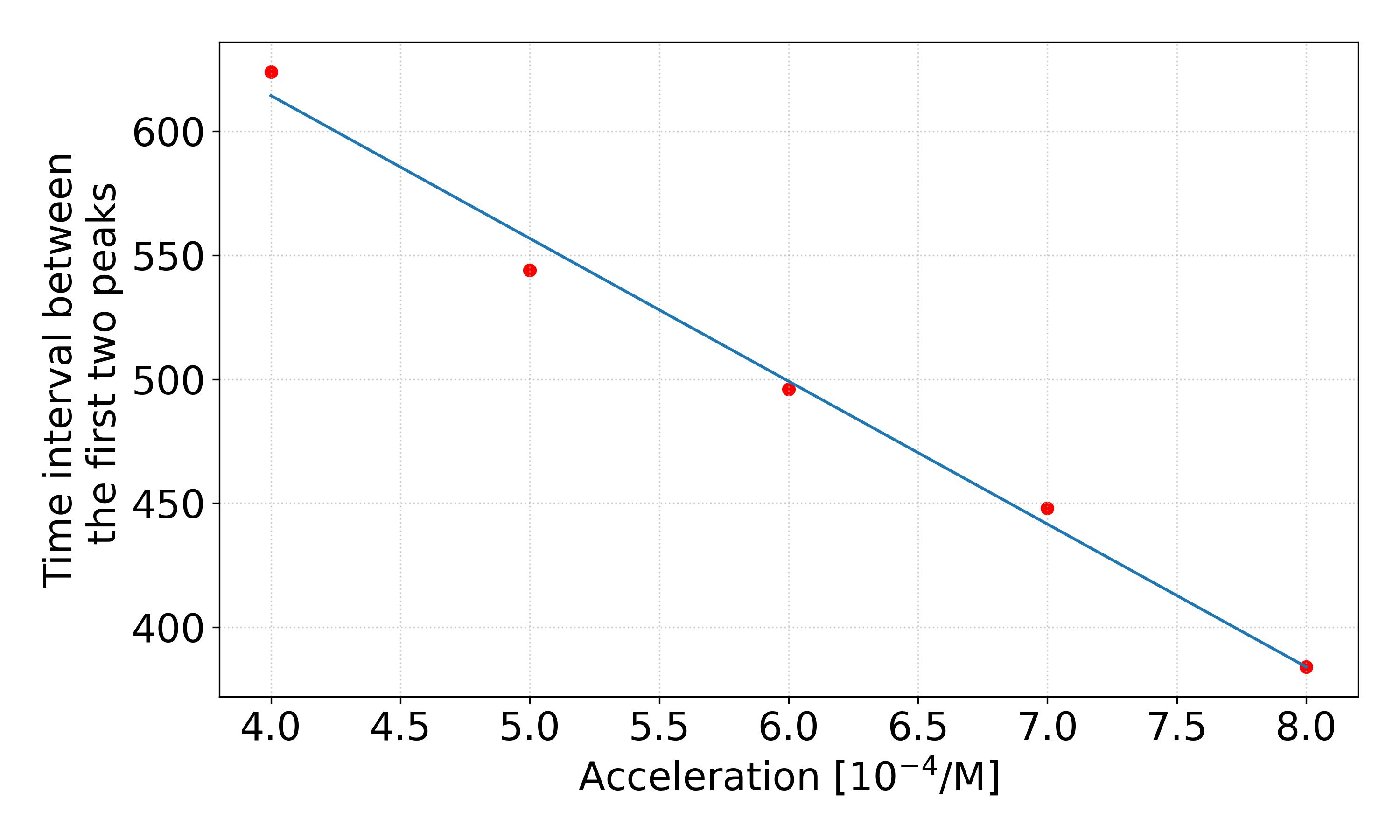}
    \caption{Variation of the time period between the detachment of the first scalar field cloud and formation of the second cloud, with acceleration, for the C-metric.}
    \label{fig:cloud_12}
\end{figure}

\begin{figure}[htbp]
    \centering
    \includegraphics[width=1\linewidth]{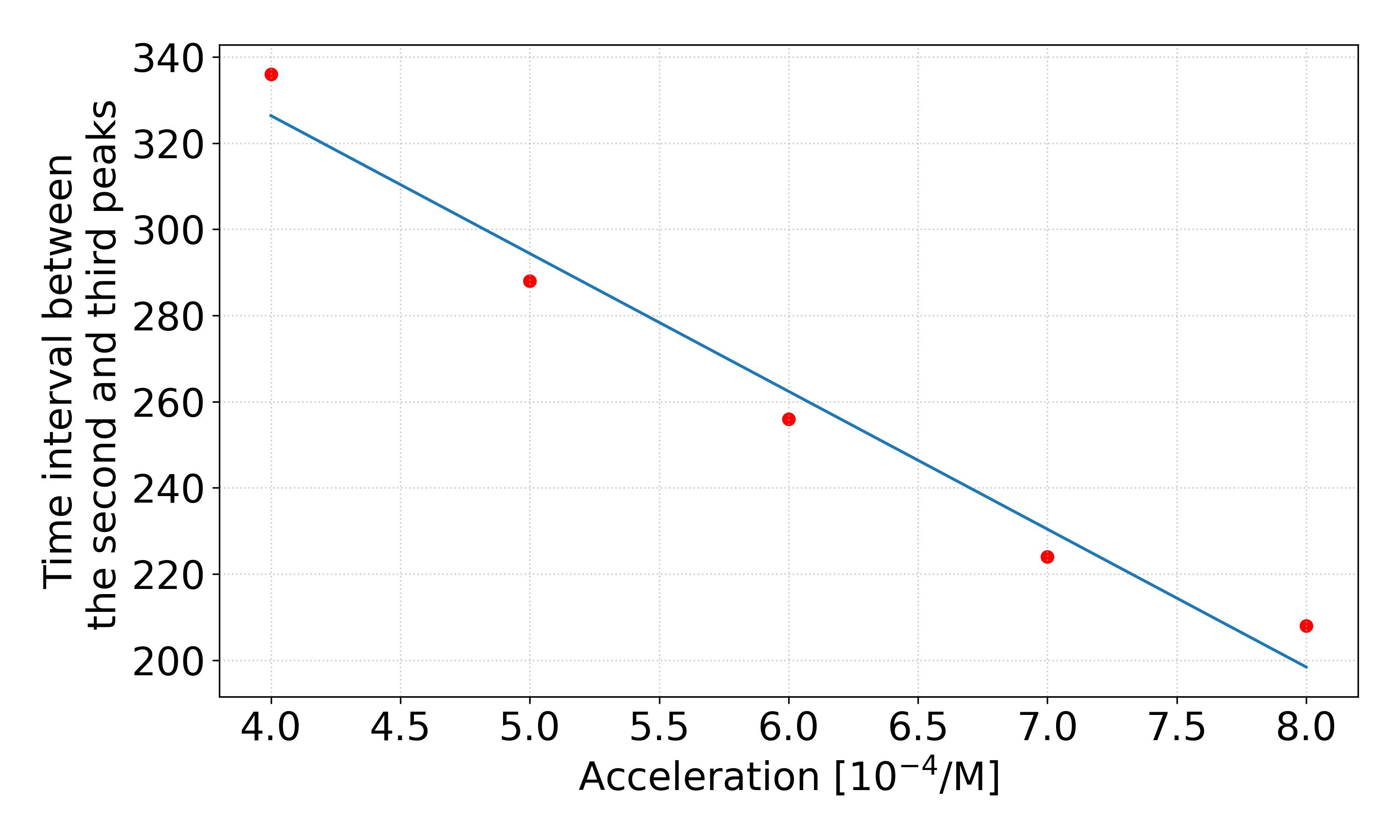}
    \caption{Variation of the time period between the detachment of the second scalar field cloud and formation of the third cloud, with acceleration, for the C-metric.}
    \label{fig:cloud_23}
\end{figure}

We also observe that the time period between the detachment of a cloud and formation of the next cloud, $t_{\rm cloud}$, reduces with simulation time $t$, as seen in figure \ref{fig:time_period_vs_t}. The inverse of the time period, with some notable exceptions, appears to follow a linear behavior with time. This is shown in figure \ref{fig:inv_time_period}. Furthermore, the time period between the detachment of first cloud and formation of second cloud, has a linear relationship with acceleration. This is also observed for the time period between the second and the third clouds. These can be seen in figures \ref{fig:cloud_12} and \ref{fig:cloud_23}. However, the slopes of these two lines are different. This shows us that the time period between two successive clouds depends inversely on the acceleration $a$ and simulation time $t$.

\begin{figure}[htbp]
    \centering
    \includegraphics[width=1\linewidth]{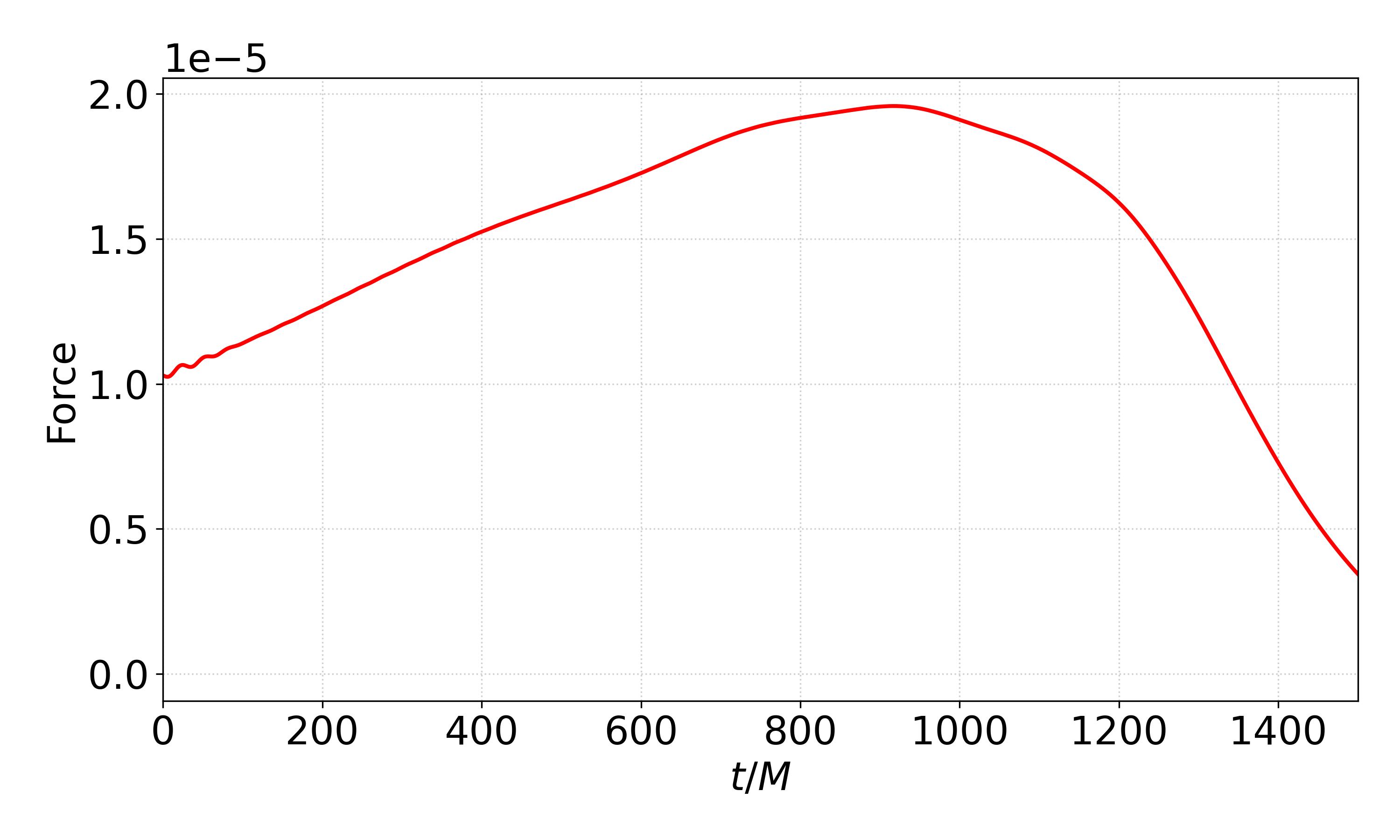}
    \caption{Variation of force on the BH($F_i$) with time for C-metric with $a = 0.0008$.}
    \label{fig:fvta}
\end{figure}

\begin{figure}[htbp]
    \centering
    \includegraphics[width=1\linewidth]{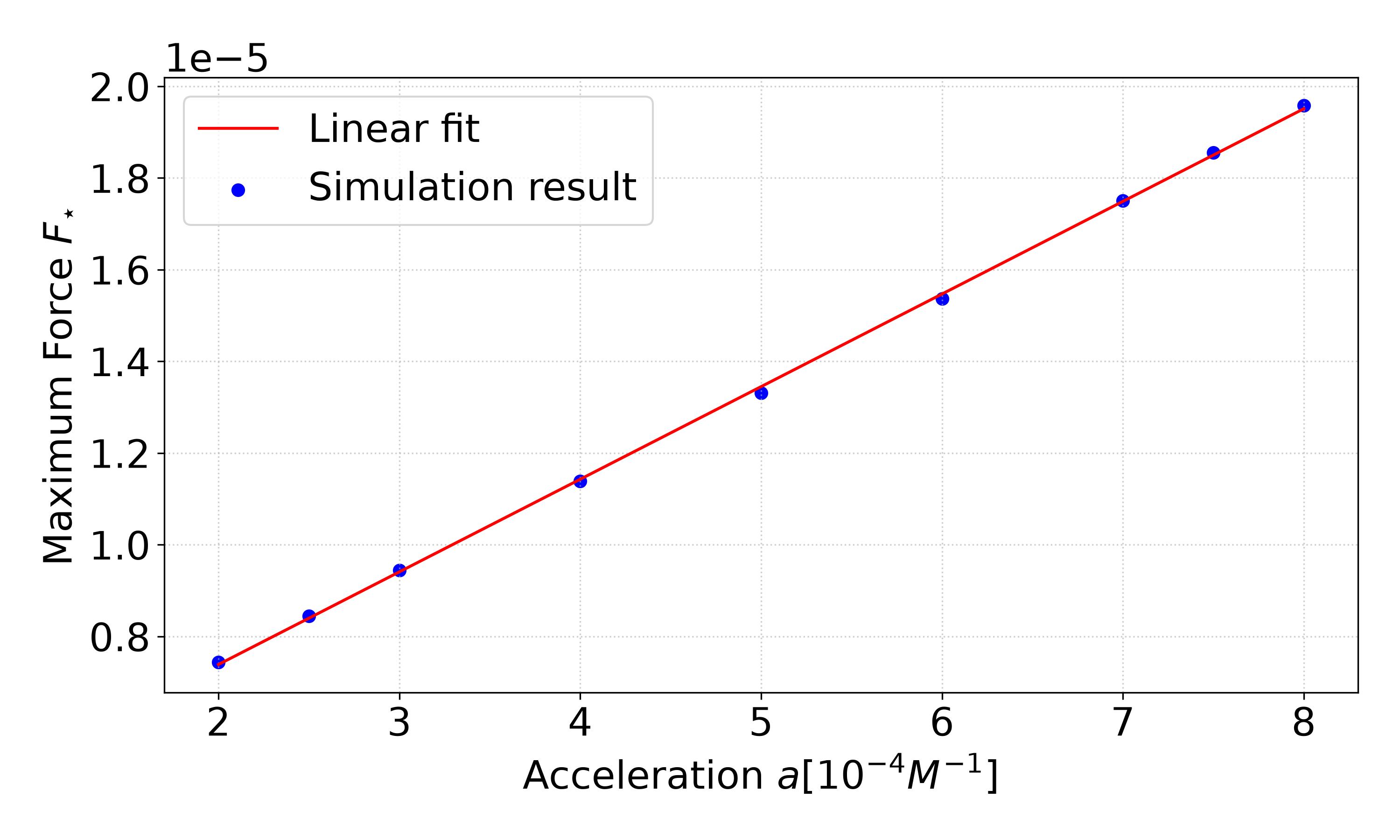}
    \caption{Variation of maximum force($F_{\star}$) with acceleration $a$ for C-metric. The slope of the linear fit is $2.02062981 \times 10^{-2}~M$ and the intercept is $3.34559047 \times 10^{-6}$.}
    \label{fig:fva}
\end{figure}

\begin{figure}[htbp]
    \centering
    \includegraphics[width=1\linewidth]{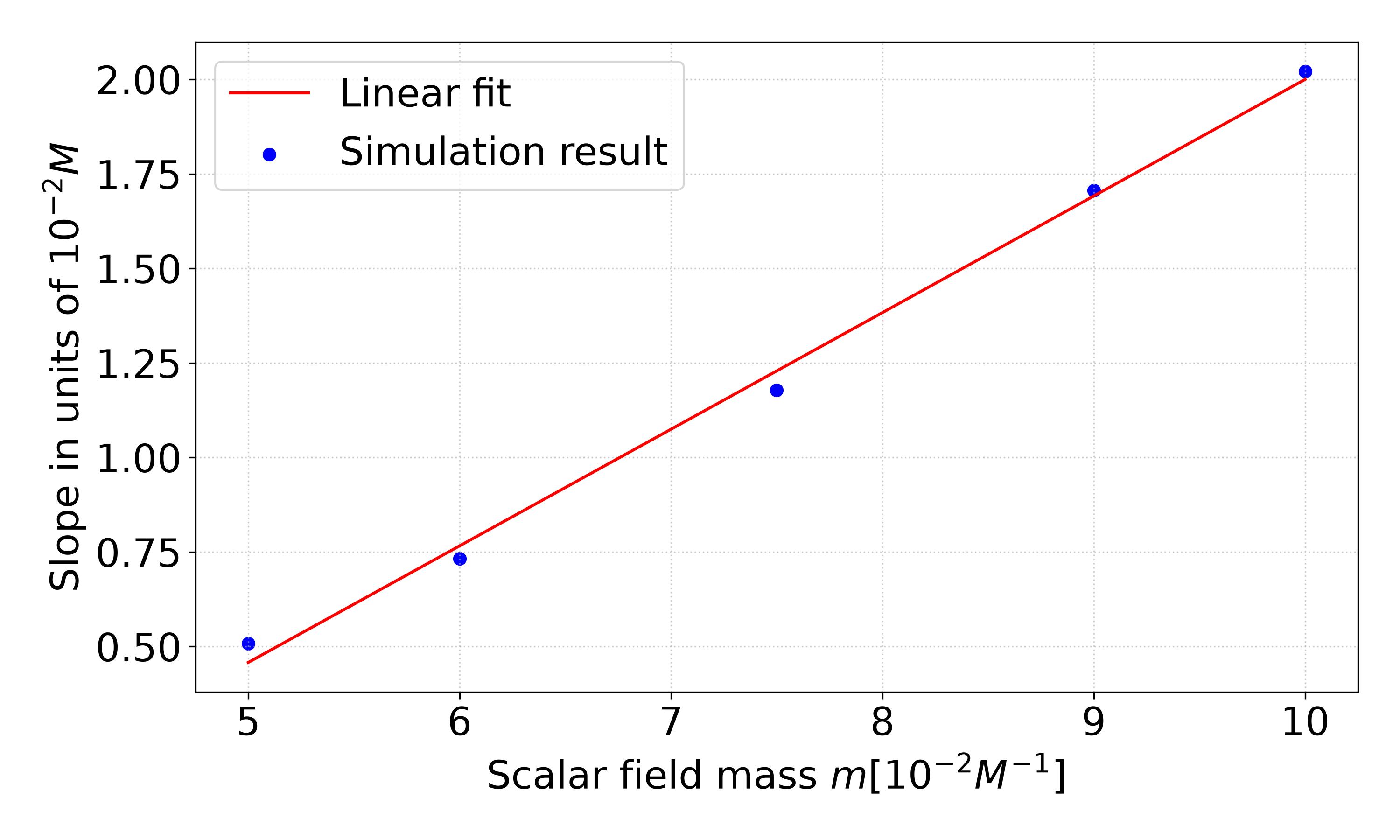}
    \caption{Variation of slope of maximum force v/s acceleration, with acceleration  $a$ for C-metric. The slope of the linear fit is $0.30839147$ and the intercept is $-0.01084038~M$.}
    \label{fig:sva}
\end{figure}

\begin{figure}[htbp]
    \centering
    \includegraphics[width=1\linewidth]{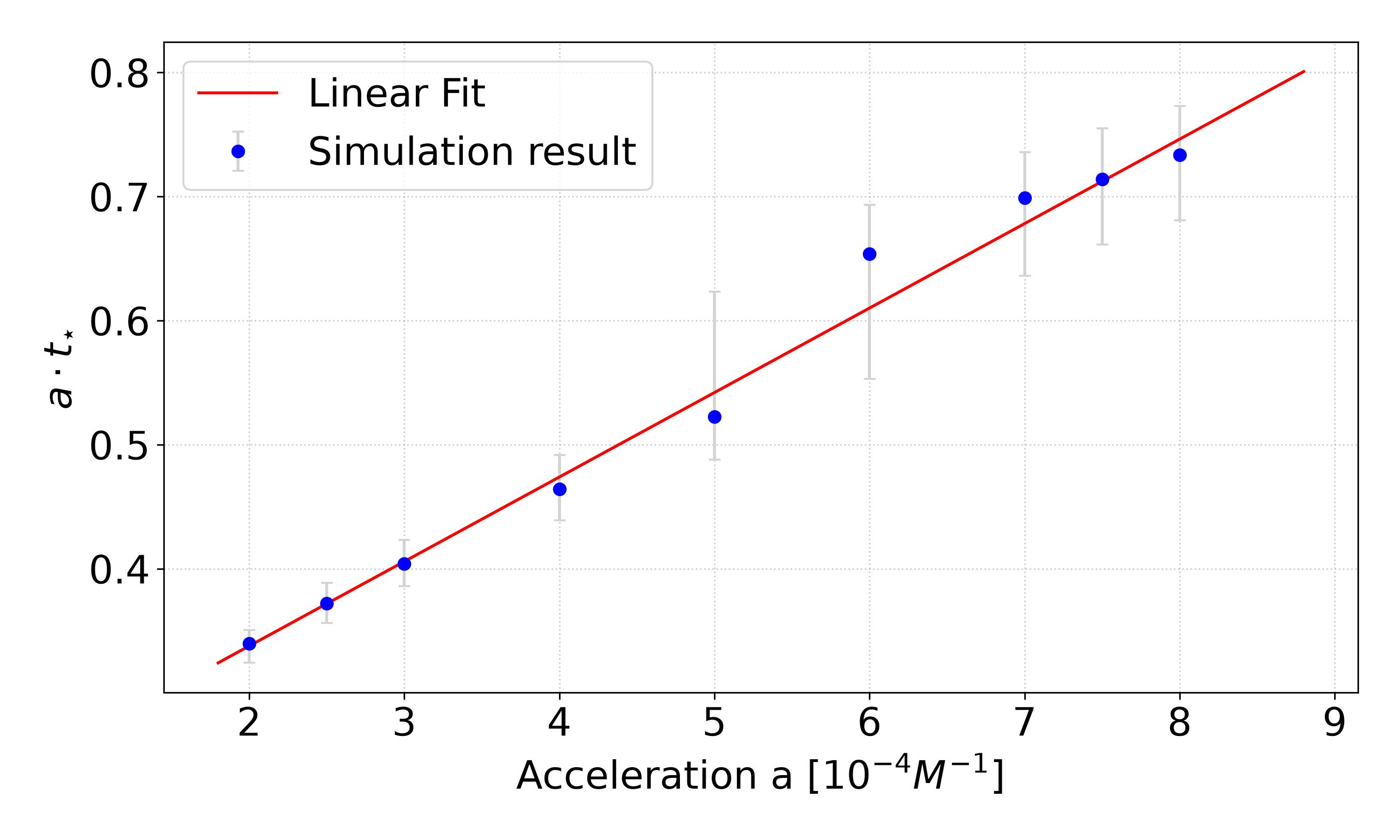}
    \caption{Variation of product of time at which force reaches maxima and acceleration $a$, with acceleration $a$ for C-metric. The error bar is incorporated because the force hovers around the maximum value for a duration of time, as seen in figure \ref{fig:fva}. The $1 \sigma$ error bar region is chosen as the time period for which force $F_z \geq 0.99 F_{\star}$. The slope is $680.88 \pm 51.15~M$ and intercept is $0.2015 \pm 0.0176$.}
    \label{fig:tva}
\end{figure}

Another interesting thing to note is that the force never saturates as seen in figure \ref{fig:fvta}, as opposed to what we see for the constant velocity case. The force seems to reach a maxima and then reduces sharply. We hypothesize that this, along with the cloud formation described in the previous paragraph, occurs because the velocity of the BH continues to increase, and the scalar field clouds cannot catch up to the high speed. This causes them to drift away and thus their force on the BH reduces. After a certain limit, catching up to the BH becomes  very hard and we then dynamical friction effects drastically reduce.


We also observe that the maximum force varies linearly with the black hole acceleration, as seen in figure \ref{fig:fva}. This suggests that there is an effect of inertia. Furthermore, the slope of the graph between maximum force and acceleration has a linear dependence on scalar field mass, as seen in figure \ref{fig:sva}. This suggests that the source of the inertia would come from mass of the scalar field, $m$. Thus we have the following dependence for maximum force $F_{\star}$,
\begin{equation}
    F_{\star} \propto m a.
\end{equation}

Furthermore, we observe that there is an inverse relationship between the time at which force reaches maxima $t_{\star}$, and acceleration. We get the following relationship,
\begin{equation}
    t_{\star} = \frac{c_1}{a} + c_2,
\end{equation}
where $c_1$ and $c_2$ are constants obtained from linear fit (see Fig. \ref{fig:tva} for more information). We further observe that the force hovers around the maximum value for a duration of time. So we have used this duration calculated as the time period for which force $F \geq 0.99 F_{\star}$ to be the $1 \sigma$ error bar for the force. We believe this feature occurs because force saturates to a maximum value and then decreases drastically because the BH is accelerating and the scalar field is unable to catch up to the BH once its velocity becomes large.

\subsection{Constant velocity BHs represented by Painlev\'e-Gullstrand  metric}

\begin{figure*}[htbp]
    \centering
    
    \subfloat[Scalar field energy density at time $t = 160 M$.]{%
        \includegraphics[width=0.48\textwidth]{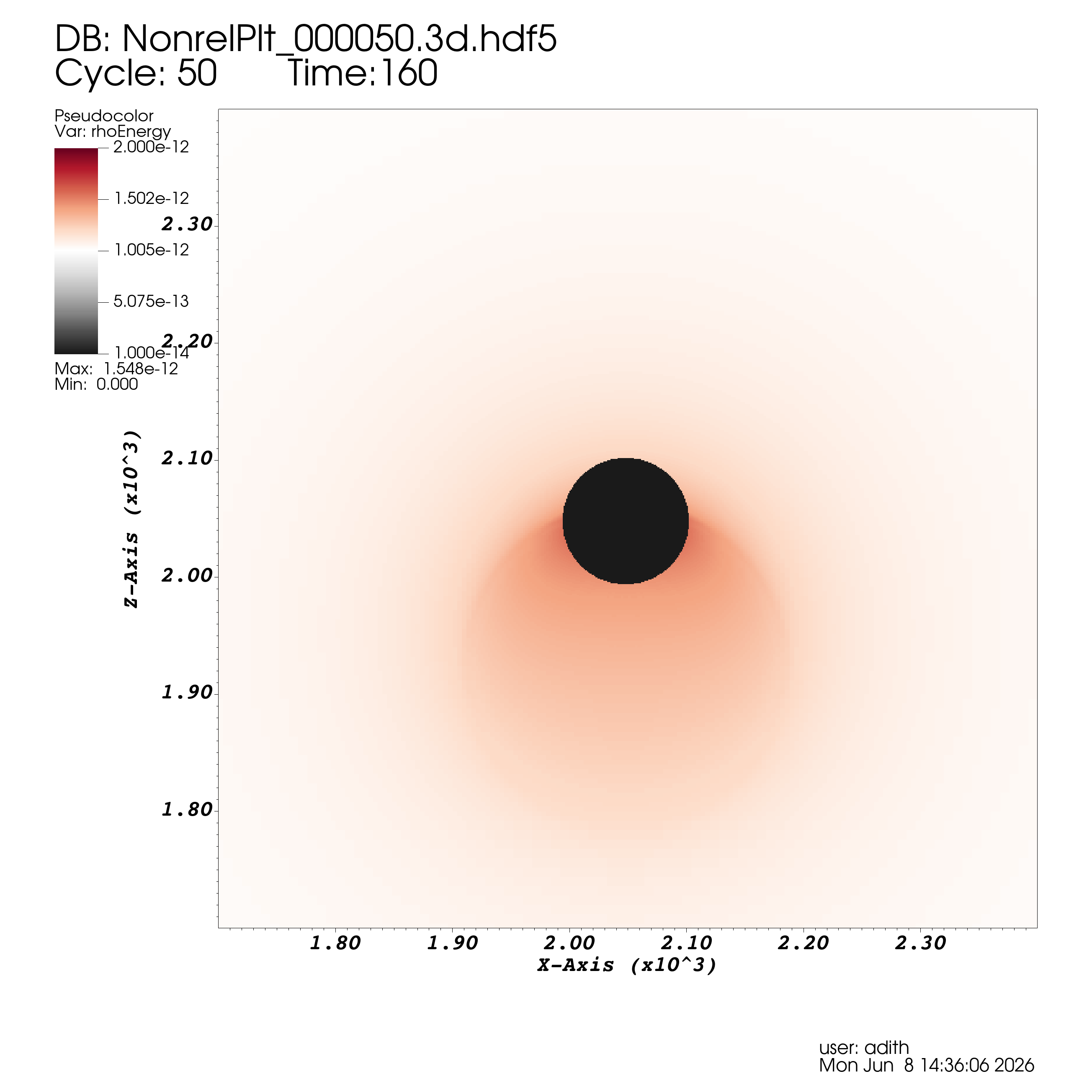}%
    }
    \hfill
    \subfloat[Scalar field energy density at time $t = 1600 M$.]{%
        \includegraphics[width=0.48\textwidth]{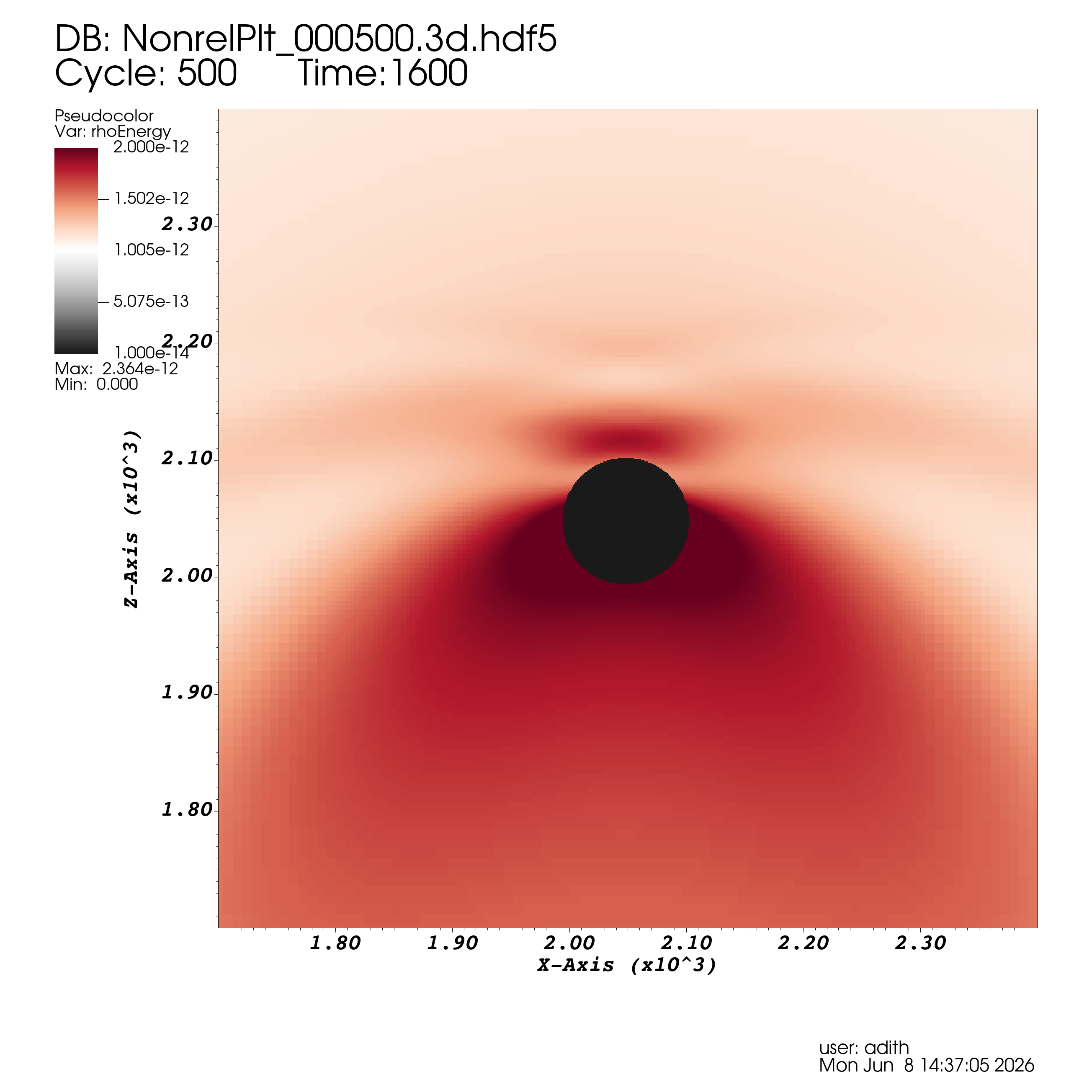}%
    }
    
    \vspace{0.5cm} 
    
    \subfloat[Scalar field energy density at time $t = 2400 M$.]{%
        \includegraphics[width=0.48\textwidth]{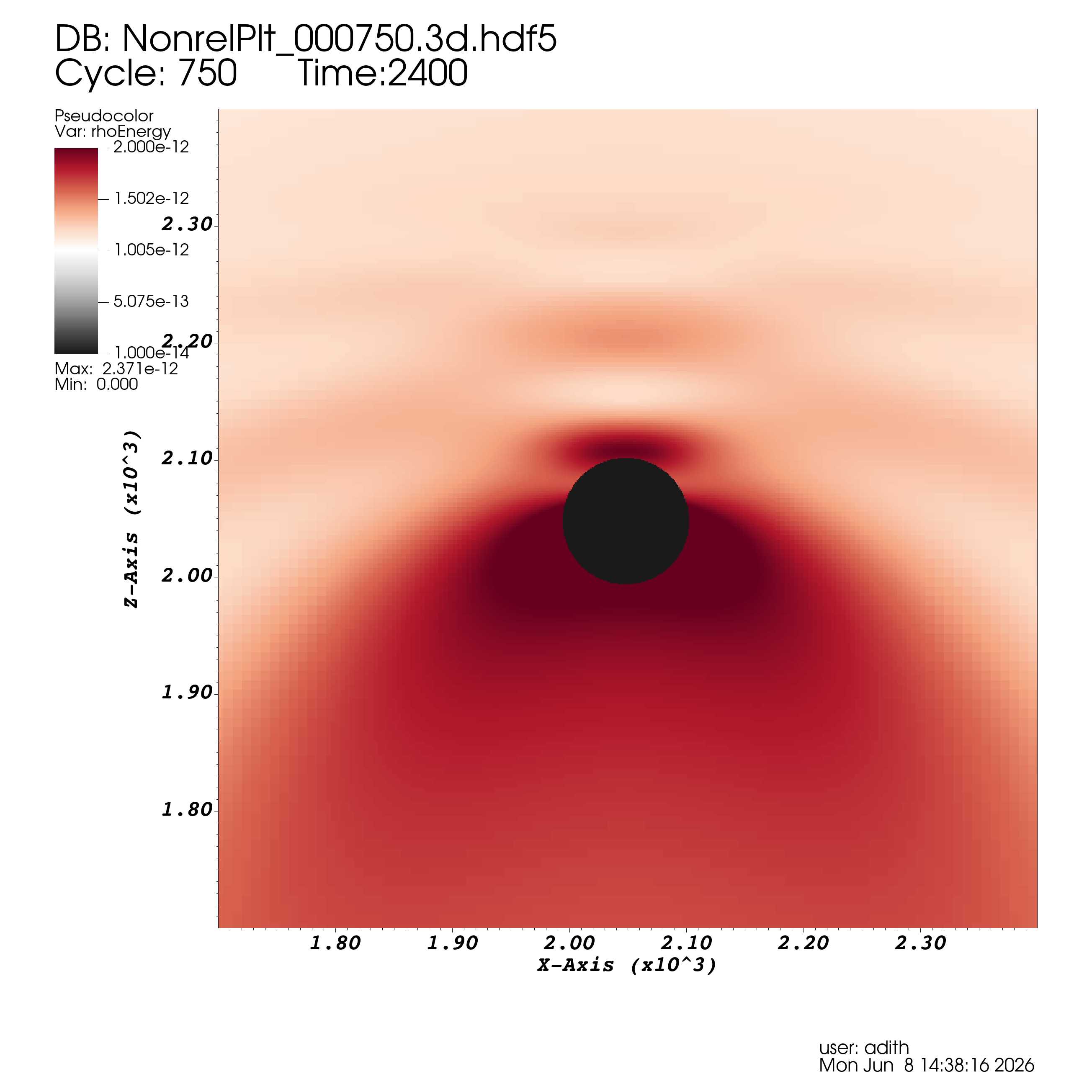}%
    }
    \hfill
    \subfloat[Scalar field energy density at time $t = 3200 M$.]{%
        \includegraphics[width=0.48\textwidth]{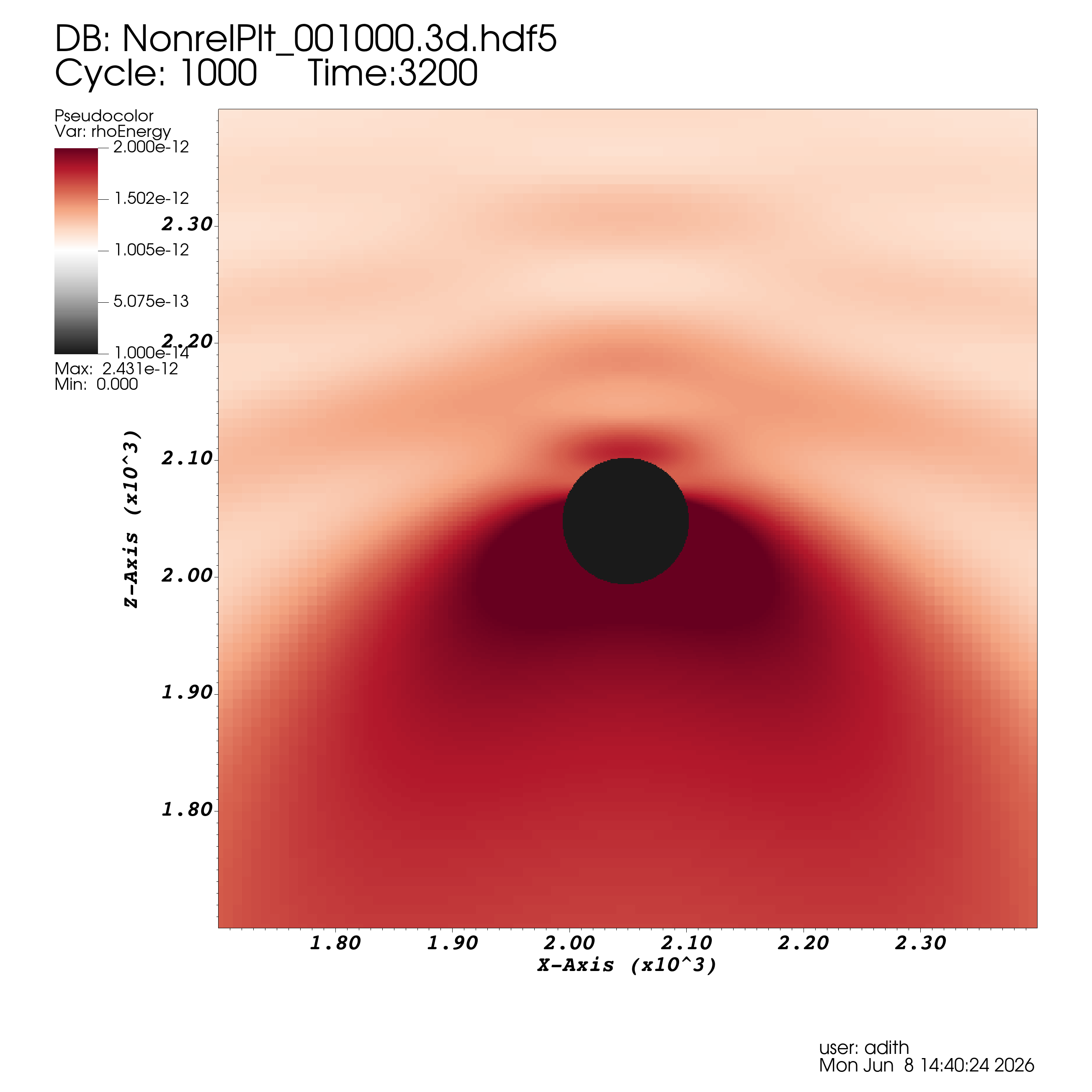}%
    }
    
    \caption{Evolution of the scalar field energy density with time for the PG metric with $v_0 = 0.8$. The distribution of energy density saturates with time. The region in black is the region not considered for force calculation. The region excised in simulations is a sphere with radius $r=1.0 M$ while the region shown in black is with $r= 70.0 M$, which encompasses the Killing horizon. This is done to ensure that the energy density $\rho$ is positive.}
    \label{fig:pg_results}
\end{figure*}

The Painlev\'e-Gullstrand  metric we have chosen represents a Schwarzschild BH moving with a constant velocity, as discussed in \ref{Painleve-Gullstrand  metric}. This represents a BH moving in the positive $z$ direction, so we expect to see energy density accumulation in the negative $z$ if dynamical friction is present. This is exactly what we observe, as seen in figure \ref{fig:pg_results}. We do not see cloud like structures forming and separating from the BH as in the C-metric case. The characteristics of the energy density are similar to what is described in~\cite{traykovadynamical2021}.

\begin{figure}[htbp]
    \centering
    \includegraphics[width=1\linewidth]{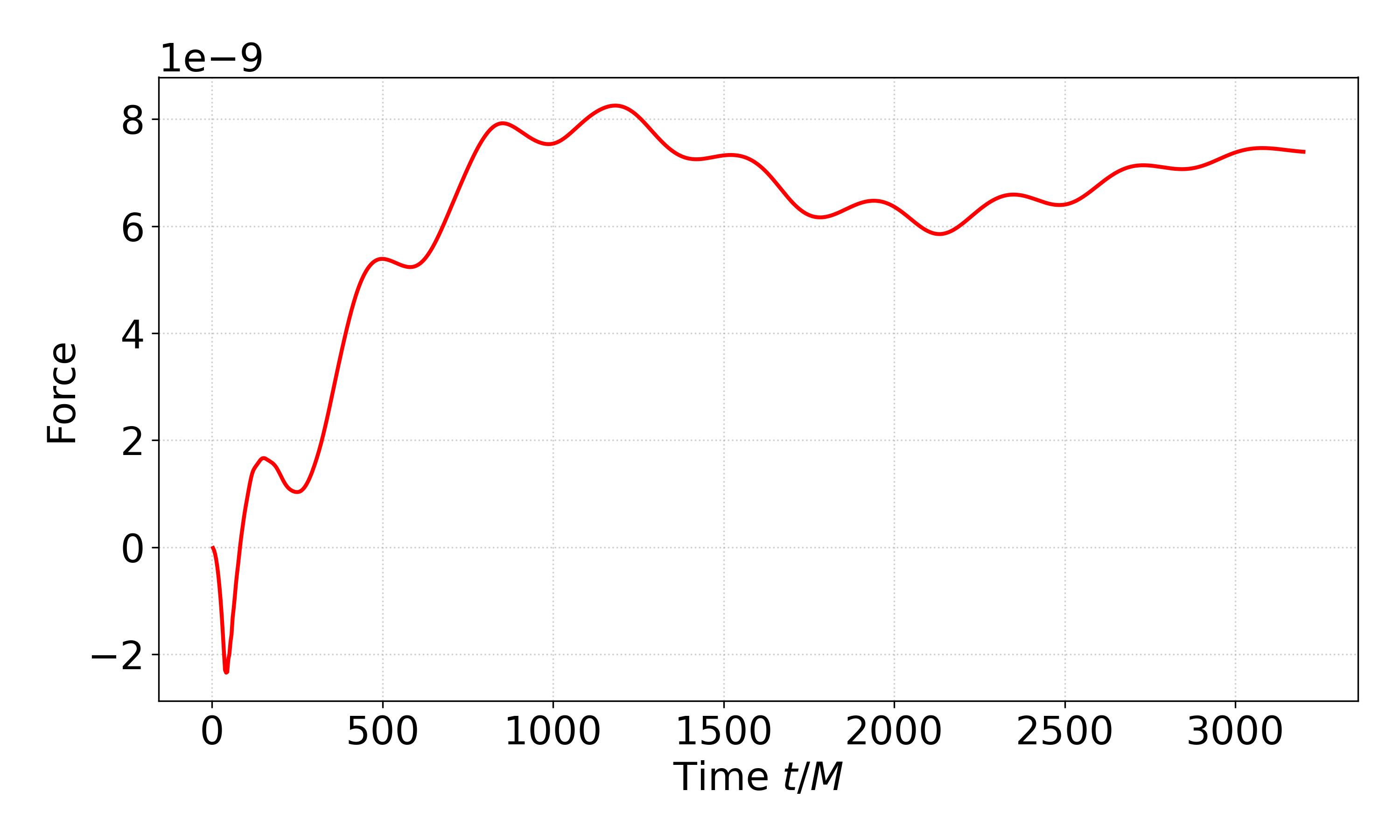}
    \caption{Variation of force on the BH in z-direction(-$F_z$) with time for PG metric with $v_0 = 0.8$.}
    \label{fig:fvtv}
\end{figure}

\begin{figure}
    \centering
    \includegraphics[width=1\linewidth]{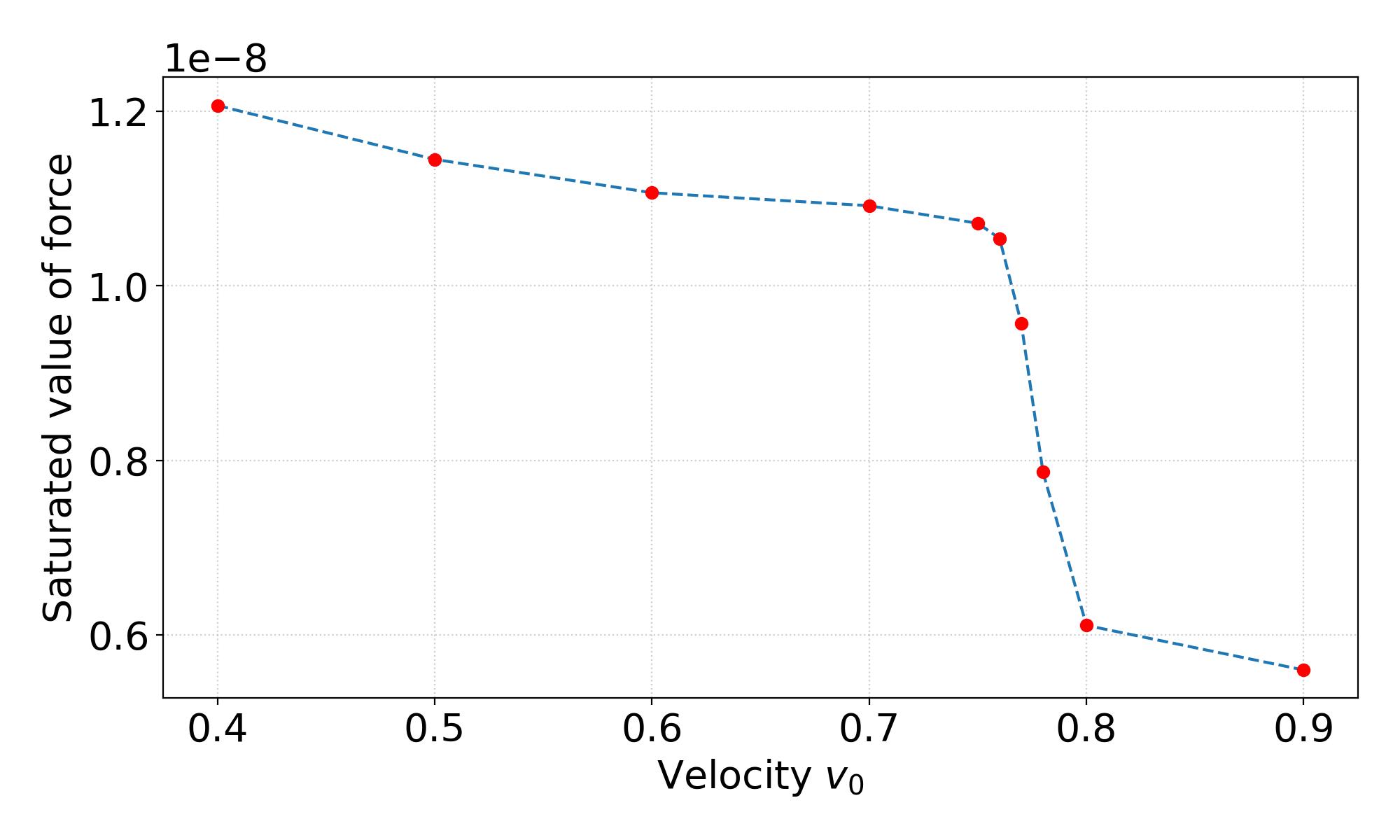}
    \caption{Variation of force at late times with velocity $v_0$ for PG metric.}
    \label{fig:fvv}
\end{figure}

One thing to note here is that the Killing horizon changes with $v_0$, and inside the Killing horizon $\rho$ becomes negative. Hence for the force calculations we only take regions outside the Killing horizon, and this varies with $v_0$. We observe saturation of force, as seen in figure \ref{fig:fvtv}. This is in accordance with what was observed in~\cite{traykovadynamical2021}.

The variation of force with velocity is presented in figure \ref{fig:fvv}. The force initially decreases with velocity gradually but then undergoes a drastic fall off around $v_0 = 0.75 \text{ to } 0.8$. This curve we observed is in contrast to what was obtained in~\cite{traykovadynamical2021}. This we expect is due to the different choice of metric and time parameter. We here choose the PG time, which is the time of a radially in-falling observer for Schwarzschild. But in~\cite{traykovadynamical2021}, they have gone with the time associated with the isotropic coordinates.

\section{Conclusion}
\label{Conclusion}

In this work, we have investigated the forces acting on BH moving through a scalar field environment, paying special attention to the distortion in spacetime geometry resulting from BH motion. Towards this end, we have considered two cases, namely: BH moving with constant acceleration, modeled by the C-metric, and BH moving with constant velocity, modeled by Painlev\'e-Gullstrand metric with a specific velocity profile. We observe the occurrence of dynamical friction in both cases, but the distribution of energy density with time and the nature of the forces varies drastically between the two case, and reveals several novel features in both the cases.

For the case of accelerating BHs, we observe the repeated formation of cloud like structures near to the BH which detach from the BH as time progresses. We observe that the time period between the detachment of one cloud and formation of the next cloud follows an inverse relationship with acceleration as well as simulation time. For the force itself, we observe that it initially increases with time and then starts to drastically decrease. We think these are because the velocity of the BH continues to increase with time, making it harder for the scalar field to catch up. This ultimately leads to a reduction in the dynamical friction effects with time. The maximum value of the force, $F_\star$, itself fits a linear relationship with acceleration, with the slope itself proportional to the scalar field mass; that is, $F_\star \propto m a$, revealing an interesting inertial like aspect of the force. While we do not have a clear, physical understanding of this result, we would like to speculate that such a proportionality is highly non-trivial and future analytical work it will motivate will lead to important insights into the fundamental nature of dynamical friction. Finally, we also observe that the time $t_\star$ at which the force reaches maxima has an inverse relationship with acceleration. 

For the constant velocity case modeled through a specific PG metric, we observe accumulation of energy density behind the BH, and also the saturation of the corresponding force, as observed in previous works~\cite{traykovadynamical2021}. However unlike the previous works, we see that the saturated value of force initially decreases gradually with velocity and then undergoes a drastic reduction around $v_0 = 0.75 \text{ to }0.8$. We expect this discrepancy is because of the difference in the metric, specifically it's non-vacuum nature, as well as the choice of PG gauge for setting up the numerical evolution case. 

We may summarize the key picture that emerges from the present work is as follows: while dynamical friction is a generic feature associated with BHs moving through an external environment, the specific details of this force depends quite sensitively on (i) how the BH motion modifies the spacetime geometry, (ii) choice of gauge used to initialize the ADM evolution data numerically, (iii) whether the BH solution being considered is a vacuum solution in absence of backreaction from the environment. Indeed, more complete numerical relativity simulations, especially for the C-metric in regimes where $\rho M^2>1$, are needed for understanding how the force changes when the backreaction from the field is non-negligible.

\clearpage

\bibliographystyle{unsrtnat}

\bibliography{ref}

\begin{thebibliography}{16}
\providecommand{\natexlab}[1]{#1}
\providecommand{\url}[1]{\texttt{#1}}
\expandafter\ifx\csname urlstyle\endcsname\relax
  \providecommand{\doi}[1]{doi: #1}\else
  \providecommand{\doi}{doi: \begingroup \urlstyle{rm}\Url}\fi

\bibitem[et~al(2016)]{gw}
B.~P.~Abbott et~al.
\newblock Observation of gravitational waves from a binary black hole merger.
\newblock \emph{Phys. Rev. Lett.}, 116:\penalty0 061102, Feb 2016.
\newblock \doi{10.1103/PhysRevLett.116.061102}.
\newblock URL \url{https://link.aps.org/doi/10.1103/PhysRevLett.116.061102}.

\bibitem[Chandrasekhar()]{chandrasekhardynamical1943}
S.~Chandrasekhar.
\newblock Dynamical friction. i. general considerations: the coefficient of dynamical friction.
\newblock 97:\penalty0 255.
\newblock ISSN 0004-637X, 1538-4357.
\newblock \doi{10.1086/144517}.
\newblock URL \url{http://adsabs.harvard.edu/doi/10.1086/144517}.

\bibitem[Traykova et~al.()Traykova, Clough, Helfer, Berti, Ferreira, and Hui]{traykovadynamical2021}
Dina Traykova, Katy Clough, Thomas Helfer, Emanuele Berti, Pedro~G. Ferreira, and Lam Hui.
\newblock Dynamical friction from scalar dark matter in the relativistic regime.
\newblock 104\penalty0 (10):\penalty0 103014.
\newblock ISSN 2470-0010, 2470-0029.
\newblock \doi{10.1103/PhysRevD.104.103014}.
\newblock URL \url{http://arxiv.org/abs/2106.08280}.

\bibitem[Dyson et~al.()Dyson, Redondo-Yuste, Meent, and Cardoso]{dysonrelativistic2024}
Conor Dyson, Jaime Redondo-Yuste, Maarten van~de Meent, and Vitor Cardoso.
\newblock Relativistic aerodynamics of spinning black holes.
\newblock 109\penalty0 (10):\penalty0 104038.
\newblock ISSN 2470-0010, 2470-0029.
\newblock \doi{10.1103/PhysRevD.109.104038}.
\newblock URL \url{http://arxiv.org/abs/2402.07981}.

\bibitem[Wang et~al.()Wang, Helfer, Traykova, Clough, and Berti]{wanggravitational2024}
Zipeng Wang, Thomas Helfer, Dina Traykova, Katy Clough, and Emanuele Berti.
\newblock Gravitational magnus effect from scalar dark matter.
\newblock URL \url{http://arxiv.org/abs/2402.07977}.

\bibitem[Griffiths et~al.()Griffiths, Krtous, and Podolsky]{griffithsinterpreting2006}
J.~B. Griffiths, P.~Krtous, and J.~Podolsky.
\newblock Interpreting the c-metric.
\newblock 23\penalty0 (23):\penalty0 6745--6766.
\newblock ISSN 0264-9381, 1361-6382.
\newblock \doi{10.1088/0264-9381/23/23/008}.
\newblock URL \url{http://arxiv.org/abs/gr-qc/0609056}.

\bibitem[Kinnersley and Walker(1970)]{KinnCM}
William Kinnersley and Martin Walker.
\newblock Uniformly accelerating charged mass in general relativity.
\newblock \emph{Phys. Rev. D}, 2:\penalty0 1359--1370, Oct 1970.
\newblock \doi{10.1103/PhysRevD.2.1359}.
\newblock URL \url{https://link.aps.org/doi/10.1103/PhysRevD.2.1359}.

\bibitem[Hong and Teo()]{hongnew2003}
Kenneth Hong and Edward Teo.
\newblock A new form of the c-metric.
\newblock 20\penalty0 (14):\penalty0 3269--3277.
\newblock ISSN 0264-9381, 1361-6382.
\newblock \doi{10.1088/0264-9381/20/14/321}.
\newblock URL \url{https://iopscience.iop.org/article/10.1088/0264-9381/20/14/321}.

\bibitem[Dray()]{drayasymptotic1982}
Tevian Dray.
\newblock On the asymptotic flatness of the c metrics at spatial infinity.
\newblock 14\penalty0 (2):\penalty0 109--112.
\newblock ISSN 0001-7701, 1572-9532.
\newblock \doi{10.1007/BF00756916}.
\newblock URL \url{http://link.springer.com/10.1007/BF00756916}.

\bibitem[Ashtekar and Dray()]{ashtekarexistence1981}
Abhay Ashtekar and Tevian Dray.
\newblock On the existence of solutions to einstein's equation with non-zero bondi news.
\newblock 79\penalty0 (4):\penalty0 581--599.
\newblock ISSN 0010-3616, 1432-0916.
\newblock \doi{10.1007/BF01209313}.
\newblock URL \url{http://link.springer.com/10.1007/BF01209313}.

\bibitem[Hamilton and Lisle()]{hamiltonriver2008}
Andrew J.~S. Hamilton and Jason~P. Lisle.
\newblock The river model of black holes.
\newblock 76\penalty0 (6):\penalty0 519--532.
\newblock ISSN 0002-9505.
\newblock \doi{10.1119/1.2830526}.
\newblock URL \url{https://doi.org/10.1119/1.2830526}.
\newblock \_eprint: https://pubs.aip.org/aapt/ajp/article-pdf/76/6/519/13137116/519\_1\_online.pdf.

\bibitem[Fischer and Visser()]{fischerspace-time2003}
Uwe~R. Fischer and Matt Visser.
\newblock On the space-time curvature experienced by quasiparticle excitations in the painleve-gullstrand effective geometry.
\newblock 304\penalty0 (1):\penalty0 22--39.
\newblock ISSN 00034916.
\newblock \doi{10.1016/S0003-4916(03)00011-3}.
\newblock URL \url{http://arxiv.org/abs/cond-mat/0205139}.

\bibitem[Aurrekoetxea et~al.(2023)Aurrekoetxea, Bamber, Brady, Clough, Helfer, Marsden, Traykova, and Wang]{Aurrekoetxea:2023fhl}
Josu~C. Aurrekoetxea, Jamie Bamber, Sam~E. Brady, Katy Clough, Thomas Helfer, James Marsden, Dina Traykova, and Zipeng Wang.
\newblock {GRDzhadzha: A code for evolving relativistic matter on analytic metric backgrounds}.
\newblock 8 2023.

\bibitem[Andrade et~al.(2021)Andrade, Salo, Aurrekoetxea, Bamber, Clough, Croft, de~Jong, Drew, Duran, Ferreira, Figueras, Finkel, Fran\c{c}a, Ge, Gu, Helfer, Jäykkä, Joana, Kunesch, Kornet, Lim, Muia, Nazari, Radia, Ripley, Shellard, Sperhake, Traykova, Tunyasuvunakool, Wang, Widdicombe, and Wong]{Andrade2021}
Tomas Andrade, Llibert~Areste Salo, Josu~C. Aurrekoetxea, Jamie Bamber, Katy Clough, Robin Croft, Eloy de~Jong, Amelia Drew, Alejandro Duran, Pedro~G. Ferreira, Pau Figueras, Hal Finkel, Tiago Fran\c{c}a, Bo-Xuan Ge, Chenxia Gu, Thomas Helfer, Juha Jäykkä, Cristian Joana, Markus Kunesch, Kacper Kornet, Eugene~A. Lim, Francesco Muia, Zainab Nazari, Miren Radia, Justin Ripley, Paul Shellard, Ulrich Sperhake, Dina Traykova, Saran Tunyasuvunakool, Zipeng Wang, James~Y. Widdicombe, and Kaze Wong.
\newblock Grchombo: An adaptable numerical relativity code for fundamental physics.
\newblock \emph{Journal of Open Source Software}, 6\penalty0 (68):\penalty0 3703, 2021.
\newblock \doi{10.21105/joss.03703}.
\newblock URL \url{https://doi.org/10.21105/joss.03703}.

\bibitem[Clough()]{cloughcontinuity2021}
Katy Clough.
\newblock Continuity equations for general matter: applications in numerical relativity.
\newblock 38\penalty0 (16):\penalty0 167001.
\newblock ISSN 0264-9381, 1361-6382.
\newblock \doi{10.1088/1361-6382/ac10ee}.
\newblock URL \url{https://iopscience.iop.org/article/10.1088/1361-6382/ac10ee}.

\bibitem[Carneiro et~al.(2022)Carneiro, Ulhoa, and Maluf]{carneiro2022black}
FL~Carneiro, SC~Ulhoa, and JW~Maluf.
\newblock On the black hole acceleration in the c-metric space-time.
\newblock \emph{Gravitation and Cosmology}, 28\penalty0 (4):\penalty0 352--361, 2022.

\end{thebibliography}

\end{document}